\documentclass[a4paper,11pt]{article}
\usepackage{jcappub}
\usepackage[utf8]{inputenc}
\usepackage{graphicx}
\usepackage{amsmath, amssymb}
\usepackage{hyperref}
\usepackage{comment}
\usepackage{bm}
\usepackage{xcolor}
\usepackage[section]{placeins}

\graphicspath{{figuras/}{./}}

\begin{document}

\title{LISA bright sirens as a tiebreaker among cosmologies in tension}

\author[a]{M. F. A. Medeiros,}
\author[a]{F. C. Carvalho,}
\author[b]{L. L. Sales}
\affiliation[a]{Departamento de Física, Universidade do Estado do Rio Grande do Norte, 59610-210 Mossoró, RN, Brazil}
\affiliation[b]{Departamento de Física, Universidade Federal de Campina Grande, Caixa Postal 10071, 58429-900 Campina Grande, Paraíba, Brazil}
\emailAdd{mateusmedeiros@alu.uern.br}
\emailAdd{fabiocabral@uern.br}
\emailAdd{lazarolima@uern.br}

\abstract{Gravitational waves from massive black hole binaries detected by LISA will deliver luminosity distances without any astrophysical calibration, and so offer a probe of the expansion history that is independent of the cosmic distance ladder. Rather than asking how tightly LISA will constrain a single fiducial cosmology, we ask which of several mutually discrepant present-day cosmologies a future LISA dataset would render most self-consistent. We build mock catalogues of LISA bright sirens under three fiducial cosmologies drawn from datasets that participate in the current $H_0$ discrepancy (the \textit{Planck}~2018 baseline, Union3, and Pantheon+SH0ES), each catalogue carrying a self-consistent error budget from instrumental noise, weak lensing, and host peculiar velocities. We compare the posterior obtained from the simulated LISA data alone with the posterior of a real-data combination of Type~Ia supernovae, DESI baryon acoustic oscillations, cosmic chronometers and the compressed \textit{Planck}~2018 CMB, and quantify their internal tension in the $(H_0,\Omega_{\rm m})$ plane. We cross each fiducial with each of three SN~Ia compilations (Pantheon+SH0ES, Union3 and DES-Dovekie) so that the influence of the assumed cosmology is separated from that of the supernova sample. Across the nine combinations the tension spans $2.08\sigma$ up to values well beyond $5\sigma$, and the split is unambiguous: changing the fiducial is what drives that range, changing the supernova compilation by at most $1.2\sigma$. Union3 and DES-Dovekie are indistinguishable throughout, since with the absolute magnitude marginalised neither constrains $H_0$. We find the discriminating power to be strongly asymmetric. A catalogue generated in a Pantheon+SH0ES universe is inconsistent with every background we test, at a significance far in excess of $5\sigma$, and the conclusion holds for much smaller catalogues: it already exceeds $5\sigma$ for $50$ bright sirens, the upper end of early forecasts for a five-year mission, and a Fisher estimate still gives $3$ to $5\sigma$ for the $10$ to $20$ events expected from more recent population models. The \textit{Planck}~2018 and Union3 fiducials, by contrast, give $2.1\sigma$ and $3.3\sigma$ against a common background, a difference comparable to the $\pm 0.4\sigma$ scatter between noise realisations, and we show that no catalogue size separates them, because the two fiducials differ by only $0.24~\mathrm{km\,s^{-1}\,Mpc^{-1}}$ in $H_0$. LISA bright sirens are therefore a sharp arbiter of the Cepheid-calibrated scenario and, on this error budget, silent between the other two. We also show that the gravitational-wave physics supplies no prior of its own on the fiducial: catalogues built at $H_0$ from $35$ to $115~\mathrm{km\,s^{-1}\,Mpc^{-1}}$ are recovered with essentially the same fractional precision, so what rejects an implausible expansion history is the electromagnetic comparison, not the waveform.}


\maketitle

\section{Introduction}
The $\Lambda$CDM model describes the Universe successfully, from primordial nucleosynthesis to the formation of large-scale structure. In recent years, however, it has faced a growing set of observational discrepancies~\cite{di2021realm, perivolaropoulos2022challenges, perivolaropoulos2024hubble, di2025cosmoverse}. The most significant of these concerns the Hubble constant, $H_0$. Measurements of the Cosmic Microwave Background (CMB) by the \textit{Planck} satellite yield $H_0 = 67.4 \pm 0.5~\mathrm{km\,s^{-1}\,Mpc^{-1}}$~\cite{aghanim2020planck}, whereas local calibrations based on Type~Ia supernovae (SNe~Ia) anchored to Cepheid variables report values near $73~\mathrm{km\,s^{-1}\,Mpc^{-1}}$~\cite{riess2022comprehensive, riess2024jwst, riess2024jwst2}, while the calibration of SNe~Ia with the tip of the red giant branch and with J-region asymptotic giant branch stars gives $H_0 \simeq 70~\mathrm{km\,s^{-1}\,Mpc^{-1}}$~\cite{freedman2025status}. The disagreement reaches $4$--$6\sigma$, depending on the data combination, and may point to new physics, to unidentified systematics, or to biases in the cosmic distance ladder itself.

Gravitational waves (GWs) from coalescing compact binaries offer a way around this calibration problem: they deliver the luminosity distance $d_L$ directly, without any intermediate astrophysical calibration~\cite{schutz1986determining, holz2005using, chen2018two}. The LIGO--Virgo--KAGRA network opened the era of GW cosmology, most notably through the binary neutron-star event GW170817, whose electromagnetic counterpart enabled the first multi-messenger measurement of $H_0$~\cite{abbott2017gw170817, abbott2017gravitational}. When no counterpart is identified, the redshift is not directly accessible, because of the mass--redshift degeneracy in the waveform; statistical methods then marginalise over the unknown host redshift using galaxy catalogues~\cite{fishbach2019standard, soares2019first, abbott2023constraints, mastrogiovanni2024cosmology}. Current GW constraints on $H_0$ remain at the $10$--$15\%$ level, limited by the small number of well-localised events and by catalogue incompleteness at $z \gtrsim 0.5$.

The Laser Interferometer Space Antenna (LISA), scheduled for launch in the 2030s, will detect GWs in the low-frequency band (from tens of $\mu$Hz to a decihertz) inaccessible to ground-based detectors~\cite{amaro2017laser, barausse2020prospects}. In this band LISA will observe the coalescence of massive black hole binaries (MBHBs, $10^4$--$10^7\,M_\odot$) out to high redshift~\cite{arun2022new, auclair2023cosmology}. Two features make these sources valuable for cosmology~\cite{auclair2023cosmology, arun2022new}: their high redshift reach, which probes the expansion history before the onset of cosmic acceleration, and the expectation that a fraction of MBHB mergers occur in gas-rich environments and produce an electromagnetic counterpart, making them \emph{bright sirens} with a measurable redshift~\cite{tamanini2016science}.

Several studies have forecast LISA's cosmological reach~\cite{tamanini2016science, wang2022forecast}. Early work used Fisher-matrix techniques to propagate parameter uncertainties from simulated event catalogues built on astrophysical models of black-hole formation. More recent analyses added the dominant systematic errors that scatter the $d_L$--redshift relation: weak gravitational lensing~\cite{speri2021testing, wang2022forecast} and host-galaxy peculiar velocities~\cite{mukherjee2021velocity, tamanini2016science}. Bright-siren forecasts in particular now reach modified-gravity parameters and dark-energy models~\cite{belgacem2019testing, mangiagli2024massive}.

Most forecasts, however, fix a single fiducial cosmology and ask how tightly LISA will constrain its parameters. A different question becomes possible once the present data are themselves in tension: \emph{if the true cosmology is the one preferred by a given electromagnetic dataset, which of the competing scenarios will LISA favour?} Rather than treating LISA as a stand-alone ruler, we use it as a tiebreaker. We build mock LISA bright-siren catalogues under three fiducial cosmologies drawn from datasets currently in tension, namely the \textit{Planck}~2018 baseline~\cite{aghanim2020planck}, Union3~\cite{rubin2025union3}, and Pantheon+SH0ES~\cite{scolnic2022pantheon+}, and ask, in each case, how consistent the simulated LISA data are with the corresponding real-data combination.

In this paper we simulate catalogues of LISA-detectable MBHB bright sirens, adopting each of the three fiducial cosmologies above. Each catalogue self-consistently includes the three dominant distance uncertainties: instrumental noise estimated from the signal-to-noise ratio of each event with the LISA sensitivity curve~\mbox{\cite{robson2019construction}}, weak-lensing dispersion~\mbox{\cite{wang2022forecast}}, and peculiar velocities~\mbox{\cite{tamanini2016science, mukherjee2021velocity}}. We then perform a Bayesian Markov Chain Monte Carlo analysis, both for the LISA-only mock data and jointly with the corresponding real datasets (SNe~Ia, DESI baryon acoustic oscillations, cosmic chronometers, and the \textit{Planck} shift parameter). We cross each fiducial with each of three SN~Ia compilations in the background, so that the influence of the assumed cosmology can be separated from that of the supernova sample. A coverage study over $300$ noise realisations per fiducial establishes that the pipeline is unbiased; comparing the LISA-only posterior with each real-data background then measures how self-consistent each fiducial is. Two things come out of it. The choice of fiducial dominates the answer, driving nearly the whole spread across the grid, while the choice of supernova compilation moves it by at most $1.2\sigma$. And the discriminating power is asymmetric: the Pantheon+SH0ES fiducial is excluded at a significance far beyond $5\sigma$, and still at $3$ to $5\sigma$ for the $10$ to $20$ bright sirens predicted by recent population models, while the \textit{Planck}~2018 and Union3 fiducials ($2.1\sigma$ and $3.3\sigma$ against a common background) are separated by little more than the realisation-to-realisation scatter and remain so however large the catalogue grows.

The paper is organised as follows. Section~\ref{sec:LCDM} reviews the flat $\Lambda$CDM model and the luminosity distance, and sets out how the three fiducial cosmologies are defined and used. Section~\ref{sec:sirens} presents the LISA bright-siren formalism and the error budget. Section~\ref{sec:data} describes the simulated and real datasets. Section~\ref{sec:results} presents the results, and Section~\ref{sec:results:fiducial} examines how much of the answer is fixed by the choice of fiducial rather than measured, including the behaviour at deliberately unphysical fiducials. Section~\ref{sec:conclusions} summarises our conclusions.

\section{Flat $\Lambda$CDM, the luminosity distance, and the three fiducial cosmologies}
\label{sec:LCDM}

\subsection{Background cosmology and distances}
\label{sec:LCDM:background}

The flat $\Lambda$CDM model is the current standard framework for cosmology. It describes a spatially flat Universe ($\Omega_k = 0$) at critical density, filled with cold dark matter, a cosmological constant $\Lambda$, and radiation. The Hubble parameter then evolves as
\begin{equation}
H(z) = H_0 \sqrt{\Omega_{m,0}(1+z)^3 + \Omega_{r,0}(1+z)^4 + \Omega_{\Lambda,0}},
\label{eq:Hz}
\end{equation}
where $H_0$ is the Hubble constant, $\Omega_{m,0}$ the present-day matter density, $\Omega_{r,0}$ the radiation density, and $\Omega_{\Lambda,0} = 1 - \Omega_{m,0} - \Omega_{r,0}$ the dark-energy contribution. At the redshifts relevant here ($z \lesssim 3$), the radiation term is negligible.

The luminosity distance follows directly from the expansion history,
\begin{equation}
d_L(z) = (1+z)\, c \int_0^z \frac{dz'}{H(z')},
\label{eq:dL}
\end{equation}
and is the key observable linking a standard candle or siren to the underlying cosmology.

For SNe~Ia, one recovers $d_L$ from the observed distance modulus,
\begin{equation}
\mu(z) \equiv m - M = 5 \log_{10}\!\left(\frac{d_L(z)}{10~\mathrm{pc}}\right),
\label{eq:mu}
\end{equation}
where $m$ and $M$ are the apparent and absolute magnitudes. SNe~Ia are standardisable candles: empirical light-curve and colour corrections calibrate their intrinsic brightness $M$, and the run of $\mu(z)$ with redshift then constrains $H_0$ and $\Omega_{m,0}$ through Eq.~\eqref{eq:dL}. This calibration, however, rests on the cosmic distance ladder and inherits its anchor dependence. Gravitational waves avoid the ladder entirely: they fix $d_L$ from the waveform alone, independently of any astronomical distance calibration.

Throughout, the model has two free parameters, $\bm{\theta} = (H_0, \Omega_{m,0})$; the radiation density is held at its standard value and all remaining quantities are derived. Additional parameters entering individual likelihoods, namely the SN~Ia absolute magnitude $M_B$ and the sound horizon $r_d$, are nuisance parameters and are treated as described in Section~\ref{sec:data}.

\subsection{Why three fiducials, and which ones}
\label{sec:LCDM:whichfid}

A forecast must assume a cosmology in order to generate mock data. The usual choice is a single fiducial, typically the \textit{Planck} best fit, and the forecast then reports the error bars LISA would achieve around it. That procedure answers a question about precision, not about which cosmology is correct, and it is silent precisely where the present data disagree.

We therefore replace the single fiducial by three. Each is the published flat-$\Lambda$CDM cosmology of a dataset that participates in the current $H_0$ discrepancy, and each defines a self-contained \emph{scenario}: a hypothetical universe in which that dataset's preferred cosmology is the true one. The three scenarios are the \textit{Planck}~2018 CMB baseline, which anchors the high-redshift, sound-horizon-calibrated end of the discrepancy; Pantheon+SH0ES, which anchors the local, Cepheid-calibrated end; and Union3, an independent SN~Ia compilation built with a different systematics treatment, which sits between the two and is tied to neither the Cepheid ladder nor a single supernova analysis pipeline. Table~\ref{tab:fiducials} lists the adopted values. For Union3 this requires the SNe+CMB combination, since supernovae with $M_B$ marginalised do not constrain $H_0$ at all (Section~\ref{sec:priors}). These values are inputs to the analysis and no run alters them. The rest of this section discusses each choice, and Section~\ref{sec:results:fiducial} measures how much the conclusions depend on them.

We briefly describe what each scenario represents. The \textit{Planck}~2018 baseline is the sound-horizon-calibrated end of the discrepancy: it fixes the expansion history by fitting the acoustic scale at recombination and extrapolating forward, so its $H_0$ is inferred rather than measured locally, and it is the value most late-time geometric probes agree with. Pantheon+SH0ES is the opposite anchor. Its supernovae constrain the shape of $d_L(z)$, and the SH0ES Cepheid ladder supplies the absolute calibration directly in the local Universe; it is the only one of the three whose $H_0$ comes from a measurement independent of any early-Universe assumption, and it is the highest. Union3 sits between them in construction rather than merely in value: it is an independent compilation of the same class of objects as Pantheon+, reduced with a different light-curve fitter and a different treatment of selection effects, so a disagreement between the two would point to supernova systematics rather than to cosmology.

A fourth compilation, DES-Dovekie~\cite{popovic2025dovekie}, is implemented in our pipeline and appears in the analysis of Section~\ref{sec:results:fiducial}, but we do not promote it to a scenario of its own, for two reasons. First, like Union3 it marginalises $M_B$ and therefore does not determine $H_0$ (Section~\ref{sec:priors}), so its scenario would need an externally chosen Hubble constant, and that choice, rather than the supernova data, would set where the scenario sits. Second, its matter density, $\Omega_{m,0} = 0.329 \pm 0.015$ in our own fit, is close enough to the Pantheon+ value of $0.334 \pm 0.018$~\cite{brout2022pantheon} that the two would occupy nearly the same point in the plane: a fourth scenario would add a label without adding a distinct hypothesis to discriminate. We therefore keep it as a consistency check rather than as a headline case.

\begin{table}[t]
\centering
\caption{The three fiducial cosmologies at which the LISA catalogue is
simulated. Each is the published flat-$\Lambda$CDM result of a dataset that
participates in the current $H_0$ discrepancy. For \textit{Planck} and Union3 both
parameters come from the same fit. The Pantheon+SH0ES row pairs the SH0ES value of
$H_0$~\cite{riess2022comprehensive} with the Pantheon+ matter density of~\cite{brout2022pantheon};
the joint Pantheon+SH0ES fit of~\cite{brout2022pantheon} gives the same $\Omega_{m,0}$
with $H_0 = 73.6 \pm 1.1~\mathrm{km\,s^{-1}\,Mpc^{-1}}$.}
\label{tab:fiducials}
\small
\setlength{\tabcolsep}{4.5pt}
\begin{tabular}{lccl}
\hline\hline
Scenario & $H_0$ [km/s/Mpc] & $\Omega_{m,0}$ & Source \\
\hline
\textit{Planck}~2018 baseline & $67.34$ & $0.3153$ & TT,TE,EE+lowE+lensing~\cite{aghanim2020planck} \\
Union3 & $67.1$ & $0.319$ & SNe+CMB~\cite{rubin2025union3} \\
Pantheon+SH0ES & $73.04$ & $0.334$ & \cite{riess2022comprehensive, brout2022pantheon} \\
\hline
\end{tabular}
\end{table}

Two features of Table~\ref{tab:fiducials} require comment, because they set the scope of what the comparison can establish.

First, a SN~Ia compilation with the absolute magnitude $M_B$ marginalised does not measure $H_0$: the supernova Hubble diagram fixes the shape of $d_L(z)$ and therefore $\Omega_{m,0}$, but its normalisation is degenerate with $M_B$. Only Pantheon+SH0ES breaks that degeneracy, through the SH0ES Cepheid calibration. For the Union3 scenario we therefore adopt the SNe+CMB combination of~\mbox{\cite{rubin2025union3}}, $h = 0.671 \pm 0.006$ and $\Omega_{m,0} = 0.319 \pm 0.008$, taking both parameters from the same fit rather than pairing a Union3 matter density with an externally chosen $H_0$.

Second, this choice has a cost: the Union3 scenario is no longer a purely supernova-defined cosmology, since its $H_0$ carries CMB information. The alternative, the Union3 SN-only value $\Omega_{m,0} = 0.356^{+0.028}_{-0.026}$, is independent of the CMB but supplies no $H_0$ at all, and sits high enough that pairing it with any external $H_0$ would inject a large offset in $\Omega_{m,0}$, unrelated to the $H_0$ discrepancy that motivates the exercise. The same caution applies to the Pantheon+ matter density, $\Omega_{m,0} = 0.334 \pm 0.018$, which is an SN-only determination while the \textit{Planck} value comes from a full CMB fit. SN-only matter densities carry uncertainties of order $0.02$--$0.03$ and sit systematically above the CMB and BAO values, so part of the internal tension reported below originates in the $\Omega_{m,0}$ direction rather than in $H_0$. We therefore separate the two directions explicitly when presenting the results.

\subsection{How the fiducials enter the analysis}
\label{sec:LCDM:howused}

Each scenario $S$ is analysed through the same three-step protocol, so that the comparison across scenarios is symmetric by construction.

\paragraph{Step 1: mock catalogue.} The fiducial $\bm{\theta}_S = (H_0, \Omega_{m,0})_S$ of Table~\ref{tab:fiducials} converts the source redshifts into true luminosity distances through Eq.~\eqref{eq:dL}. The intrinsic population (the redshifts, the component masses, and the orientation angles of every binary) is drawn once, with a fixed random seed, and reused unchanged in all three scenarios. Only the redshift-to-distance mapping, and hence the signal-to-noise ratio and the per-event uncertainty of Section~\ref{sec:sirens}, is recomputed at $\bm{\theta}_S$. Two catalogues from different scenarios therefore contain the same binaries observed in different universes, which isolates the effect of the cosmology from the sampling noise of the population. The observational scatter is drawn from an independent seed per scenario.

\paragraph{Step 2: real-data background.} In parallel, each scenario carries a fixed combination of present-day electromagnetic measurements, $\mathcal{L}^S_{\rm bg}$, described in Section~\ref{sec:data}. The combination differs between scenarios only in which SN~Ia compilation it contains (Pantheon+SH0ES, Union3, or none in the \textit{Planck} reference scenario), while the BAO, cosmic-chronometer, and CMB components are common to all three. The background is real data and is therefore identical in every respect to the analysis one would perform today; it is not simulated at $\bm{\theta}_S$.

\paragraph{Step 3: comparison.} We sample three posteriors per scenario with the affine-invariant sampler \texttt{emcee}~\mbox{\cite{goodman2010ensemble, foreman2013emcee}}: from the mock LISA data alone, from the background alone, and from their product. The LISA-only posterior tests whether the pipeline recovers $\bm{\theta}_S$ without bias. The comparison between the LISA-only and background posteriors defines the tiebreaker statistic: it measures how far the scenario's own fiducial sits from what present electromagnetic data prefer, in units of the combined uncertainty, with the LISA mock supplying the uncertainty a future bright-siren catalogue would contribute.

The scope of this statistic should be clear. Because the mock is generated at $\bm{\theta}_S$ and the LISA-only posterior recovers $\bm{\theta}_S$ by construction, the tension we report is, to a good approximation, the distance between the assumed fiducial and the background best fit, weighted by the precision LISA would bring. The simulated data do not adjudicate between the scenarios on their own; what they contribute is the error budget that determines whether a given offset would be resolvable. The exercise is therefore a statement about the discriminating power of a future LISA bright-siren catalogue, not a measurement of which present-day cosmology is correct.

\section{Bright sirens with LISA}
\label{sec:sirens}

A massive black hole binary becomes a \emph{bright siren} when its coalescence is accompanied by an electromagnetic (EM) counterpart. The GW signal then fixes the luminosity distance $d_L$ directly, while the counterpart provides the redshift through identification of the host galaxy~\cite{tamanini2016science, mangiagli2024massive}. We assume throughout that a counterpart is detected and the redshift is known; the inference therefore uses the measured $z$ for each event.

LISA records the strain induced by a passing GW, a linear combination of the two polarisation states~\cite{sathyaprakash2009physics},
\begin{equation}
h(t) = F_+(\theta,\phi,\psi)\,h_+ + F_\times(\theta,\phi,\psi)\,h_\times,
\end{equation}
where $\theta$ and $\phi$ locate the source on the sky, $\psi$ is the polarisation angle, and $F_+$, $F_\times$ are the antenna pattern functions.

We work in the long-wavelength limit, in which LISA reduces to two independent Michelson-like channels, $A$ and $E$, equivalent to two detectors rotated by $\pi/4$ in azimuth~\cite{cutler1998angular, robson2019construction}. For the $A$ channel,
\begin{align}
F_+^{A}(\theta,\phi,\psi) &= \frac{\sqrt{3}}{2}\Big[
  \tfrac{1}{2}(1+\cos^2\theta)\cos 2\phi\,\cos 2\psi
  - \cos\theta\,\sin 2\phi\,\sin 2\psi \Big], \\
F_\times^{A}(\theta,\phi,\psi) &= \frac{\sqrt{3}}{2}\Big[
  \tfrac{1}{2}(1+\cos^2\theta)\cos 2\phi\,\sin 2\psi
  + \cos\theta\,\sin 2\phi\,\cos 2\psi \Big],
\end{align}
and the $E$-channel patterns follow from $F^{E}(\theta,\phi,\psi) = F^{A}(\theta,\phi-\pi/4,\psi)$~\cite{cutler1998angular}. In the production runs the second channel was obtained instead by the replacement $\psi \to \psi + \pi/4$, which gives the same response once averaged over the sky but not the same dependence on sky position and polarisation. We checked that the exact form raises the median signal-to-noise ratio of the catalogue by $8\%$ and changes every tension of Table~\ref{tab:tension} by less than $0.1\sigma$. The factor $\sqrt{3}/2$ reflects the $60^\circ$ opening angle of the LISA arms~\cite{cornish2003lisa}, in contrast with the $90^\circ$ of ground-based detectors.

In the stationary-phase approximation, the Fourier-domain waveform reads
\begin{equation}
\mathcal{H}(f) = \mathcal{A}\, f^{-7/6}\, e^{i\Psi(f)},
\end{equation}
with amplitude
\begin{equation}
\mathcal{A} = \frac{\mathcal{M}_c^{5/6}}{d_L}\,\pi^{-2/3}\sqrt{\frac{5}{96}}
\sqrt{\big[F_+^{(i)}(1+\cos^2\iota)\big]^2 + \big(2 F_\times^{(i)}\cos\iota\big)^2},
\end{equation}
where $\mathcal{M}_c$ is the redshifted chirp mass, $\iota$ the inclination angle, and $d_L$ the luminosity distance.

The signal-to-noise ratio (SNR) $\rho$ determines whether an event is detected. We sum in quadrature over the two sensitive channels,
\begin{equation}
\rho = \left[\big(\rho^{A}\big)^2 + \big(\rho^{E}\big)^2\right]^{1/2},
\qquad
\big(\rho^{(i)}\big)^2 = 4\int_{f_{\min}}^{f_{\max}}
\frac{|\mathcal{H}(f)|^2}{S_h(f)}\,df,
\end{equation}
with the third (null) TDI channel $T$ discarded, as it is insensitive to GWs in the relevant band. We integrate over $f \in [10^{-4},\,2\times 10^{-2}]~\mathrm{Hz}$ and treat any event with $\rho > 8$ as detected~\cite{tamanini2016science}.

For the noise power spectral density $S_h(f)$ we adopt the analytic LISA sensitivity of Robson, Cornish \& Liu~\cite{robson2019construction}, which combines the optical-metrology and test-mass acceleration noises with the confusion foreground from unresolved Galactic binaries, for an arm length $L = 2.5\times10^{9}\,\mathrm{m}$. The explicit terms are given in~\cite{robson2019construction}.

Three sources dominate the uncertainty on each measured distance. First, the instrumental error from the GW parameter estimation, which we approximate as
\begin{equation}
\sigma_{d_L}^{\rm inst} \approx \frac{2\, d_L}{\rho},
\end{equation}
where the factor of two accounts for the correlation between $d_L$ and the inclination angle, whose largest effect on the signal-to-noise ratio is a factor of two~\cite{zhao2011determination, cai2017estimating}. Second, weak gravitational lensing scatters the distance by
\begin{equation}
\sigma_{d_L}^{\rm lens} = d_L \times 0.066
\left[\frac{1-(1+z)^{-0.25}}{0.25}\right]^{1.8},
\end{equation}
following~\mbox{\cite{hirata2010reducing, tamanini2016science, wang2022forecast}}. Third, the peculiar velocity of the host galaxy contributes~\mbox{\cite{kocsis2006finding, wang2022forecast}}
\begin{equation}
\sigma_{d_L}^{\rm pec} = d_L\,\frac{\sqrt{\langle v^2\rangle}}{c}
\left[1 + \frac{c(1+z)^2}{H(z)\,d_L}\right],
\end{equation}
where we take $\sqrt{\langle v^2\rangle} = 500~\mathrm{km\,s^{-1}}$ for the typical peculiar velocity relative to the Hubble flow. The three contributions add in quadrature,
\begin{equation}
\sigma_{d_L} = \sqrt{(\sigma_{d_L}^{\rm inst})^2
+ (\sigma_{d_L}^{\rm lens})^2 + (\sigma_{d_L}^{\rm pec})^2}.
\label{eq:total}
\end{equation}

For each detected event we draw the observed distance from a Gaussian centred on the fiducial value, $d_L^{\rm obs} \sim \mathcal{N}(d_L,\sigma_{d_L})$. Because this scatter is applied after the SNR selection and does not feed back into it, the observed distances remain unbiased with respect to the fiducial cosmology; no Malmquist-type correction is therefore required in the likelihood.

\begin{figure}[!htbp]
\centering
\includegraphics[width=\linewidth]{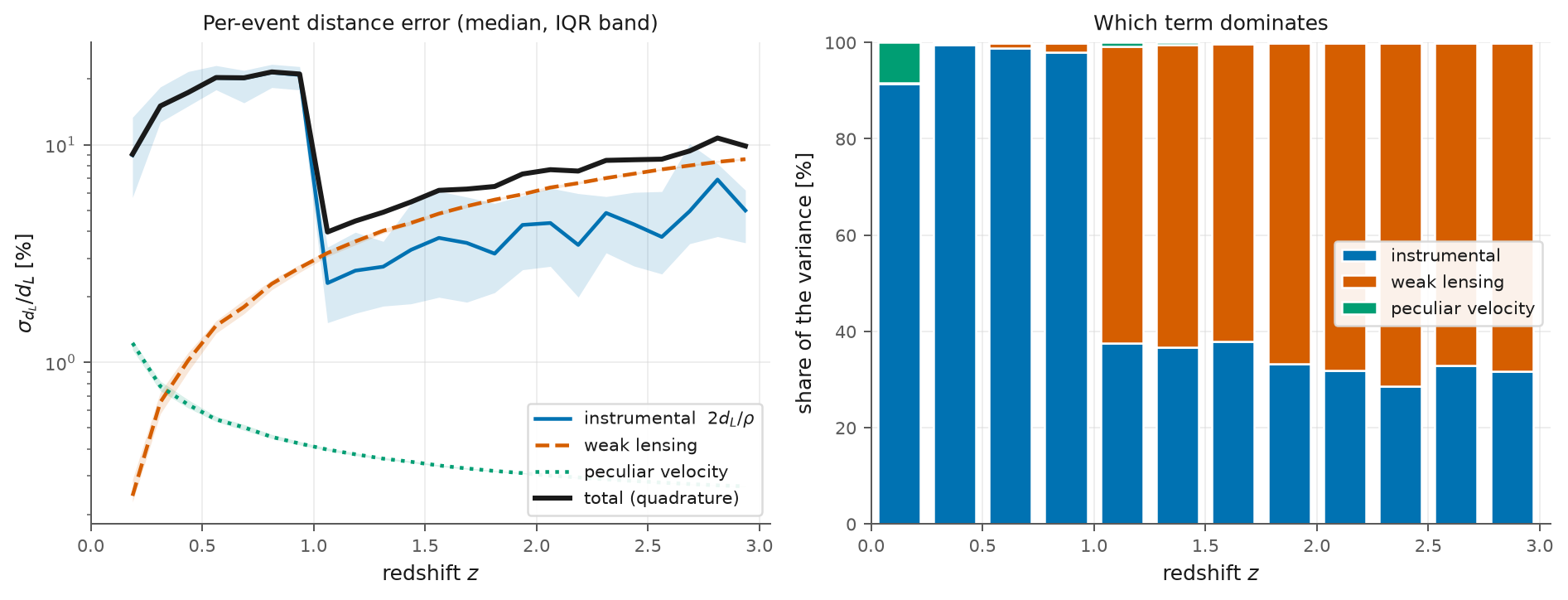}
\caption{Decomposition of the per-event distance uncertainty. \emph{Left}: median fractional error of each term against redshift, with the interquartile band; the instrumental term scatters widely because the signal-to-noise ratio depends on the drawn masses and orientation. \emph{Right}: share of the total variance carried by each term, in redshift bins. The step at $z = 1$ follows the population model, whose component-mass range changes there.}
\label{fig:budget}
\end{figure}

Figure~\ref{fig:budget} shows how the three terms combine. Weak lensing dominates the variance above $z \simeq 1$ and the instrumental term below it, while the peculiar-velocity contribution stays at the sub-percent level throughout and carries $0.2\%$ of the total variance: the catalogue contains too few low-redshift events for it to matter.

\section{Datasets}
\label{sec:data}

We test three fiducial cosmologies, each drawn from a dataset that participates in the current $H_0$ discrepancy: the \textit{Planck}~2018 baseline~\mbox{\cite{aghanim2020planck}}, Union3~\mbox{\cite{rubin2025union3}}, and Pantheon+SH0ES~\mbox{\cite{scolnic2022pantheon+}}. For each, we generate a mock LISA bright-siren catalogue at the corresponding $(H_0,\Omega_{m,0})$ of Table~\ref{tab:fiducials} and confront it with the real electromagnetic data. This section describes both the simulated and the real datasets.

\subsection{Simulated LISA catalogue}

Generating a catalogue requires, for each event, the redshift $z$, the luminosity distance $d_L(z)$, and the redshifted chirp mass $\mathcal{M}_c$. We draw redshifts from the comoving merger distribution,
\begin{equation}
\frac{dN}{dz} \propto \frac{4\pi\, d_C^2(z)\, R(z)}{H(z)\,(1+z)},
\label{eq:dNdz}
\end{equation}
where, as a phenomenological choice, we adopt the piecewise-linear merger rate of~\cite{cutler2006bbo}, a fit to the neutron-star binary rate of~\cite{schneider2001low} that was also used for ground-based sources in~\cite{nishizawa2011tracing, teixeira2023forecasts},
\begin{equation}
R(z) =
\begin{cases}
1 + 2z & \text{if } z \leq 1, \\
0.75\,(5 - z) & \text{if } 1 < z < 5, \\
0 & \text{if } z \geq 5.
\end{cases}
\label{eq:Rz}
\end{equation}
Component masses are drawn independently from a power law $p(M) \propto M^{\alpha}$ with $\alpha = -2.5$, over $[10^{2},\,10^{4}]\,M_\odot$ for $z < 1$ and $[10^{4},\,10^{6}]\,M_\odot$ for $z \geq 1$, following the observational requirements of the mission proposal for intermediate-mass black hole binaries at low redshift and massive black hole binaries at higher redshift~\cite{amaro2017laser}. The redshifted chirp mass of each binary is
\begin{equation}
\mathcal{M}_c = (1+z)\,\frac{(M_1 M_2)^{3/5}}{(M_1+M_2)^{1/5}},
\label{eq:chirp}
\end{equation}
which is the only combination of the two masses the inspiral amplitude depends on. We retain events out to $z = 3$, beyond which LISA loses sensitivity to these sources.

This population is phenomenological, and we treat it as a controlled input rather than a prediction. It should not be mistaken for a physical model of massive black hole formation, and the alternatives are neither few nor equivalent. Semi-analytic galaxy-formation models predict LISA MBHB catalogues that differ by more than an order of magnitude in event count and differ in shape as well, depending on whether black-hole seeds are light (Population~III remnants) or heavy (direct collapse) and on the delay between galaxy and black-hole merger; \mbox{\cite{klein2016science}} works through three such models and \mbox{\cite{barausse2020prospects, arun2022new}} survey the resulting spread. A parametric route is also available, with a power law in primary mass with index $\alpha$ together with a power law in mass ratio with index $\beta$, as in the framework fitted to the stellar-mass population by~\mbox{\cite{abbott2023population}}, but the parameter values measured there apply to $5$--$100\,M_\odot$ black holes and carry no information about the $10^{4}$--$10^{6}\,M_\odot$ regime. We use the simple power law above because what this work depends on is the \emph{distribution of per-event distance errors} at a given catalogue size, and we test the sensitivity of every conclusion to that size explicitly in Section~\ref{sec:results:fiducial} rather than to the population model that produced it. Section~\ref{sec:conclusions} returns to what a semi-analytic population would change.

\begin{figure*}[!htbp]
\centering
\includegraphics[width=\linewidth]{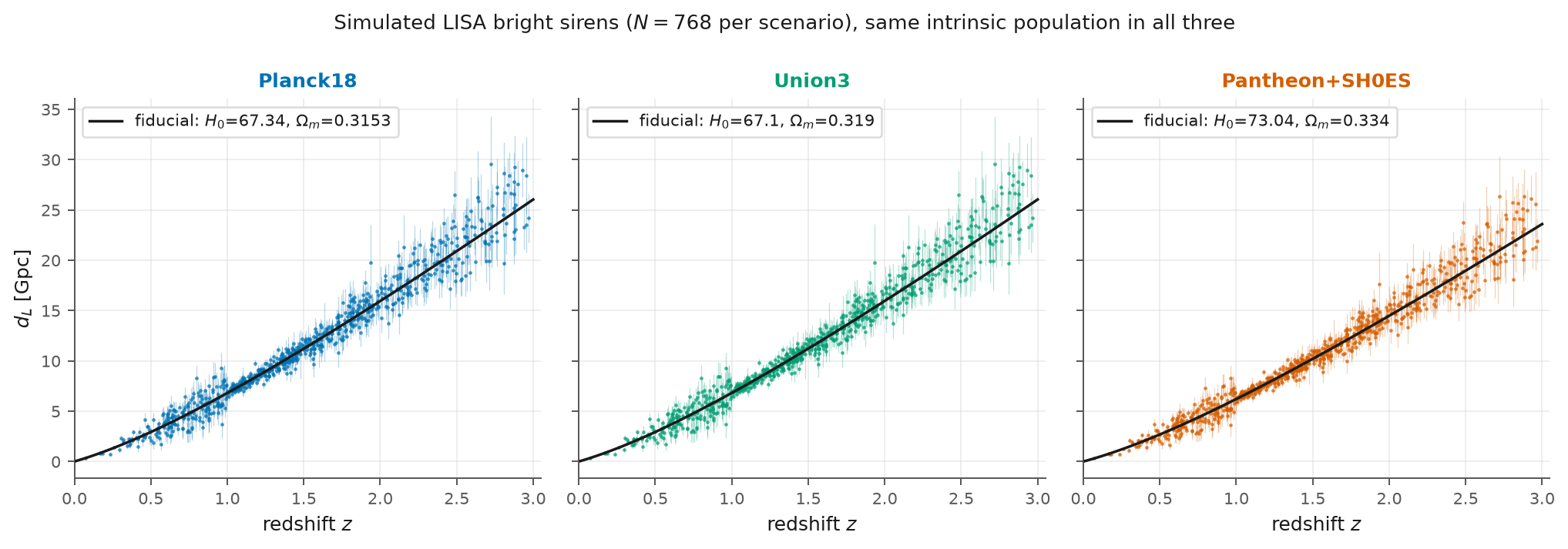}
\caption{Simulated LISA bright-siren Hubble diagrams, one per fiducial scenario. Error bars are $1\sigma$ from the combined instrumental, weak-lensing, and peculiar-velocity budget of Eq.~\eqref{eq:total}. The three panels contain the same intrinsic binaries observed in different universes: only the redshift-to-distance mapping changes, which isolates the effect of the cosmology from the sampling noise of the population.}
\label{fig:mock}
\end{figure*}

The same redshift population is used in all three scenarios; only the fiducial $(H_0,\Omega_{m,0})$ that converts redshift into distance differs. After applying the SNR selection of Sec.~\ref{sec:sirens}, we retain $N = 768$ detected bright sirens per scenario. We assign each a distance error from Eq.~\eqref{eq:total} and scatter the observed value accordingly. The resulting Hubble diagram is shown in Fig.~\ref{fig:mock}.

\begin{figure}[!htbp]
\centering
\includegraphics[width=\linewidth]{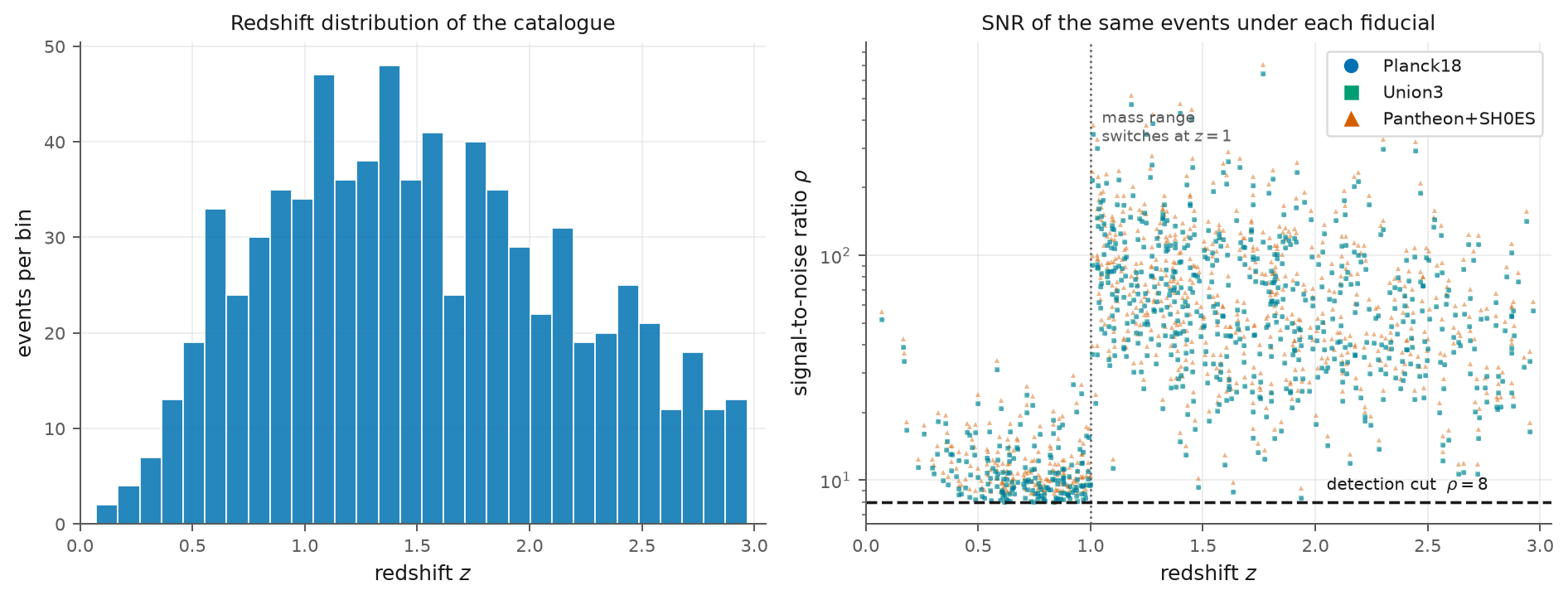}
\caption{Properties of the simulated catalogue. \emph{Left}: redshift distribution of the retained events. \emph{Right}: signal-to-noise ratio against redshift under each fiducial, with the detection threshold $\rho = 8$; the three sets overlap almost exactly because the fiducial changes only the distance scale.}
\label{fig:population}
\end{figure}

Figure~\ref{fig:population} summarises the catalogue: the redshift distribution that follows from Eq.~\eqref{eq:dNdz}, and the signal-to-noise ratio of the same events evaluated under each fiducial. Because the fiducial only rescales the redshift-to-distance mapping, the three SNR sets nearly coincide.

The Gaussian likelihood for the GW data is
\begin{equation}
\ln \mathcal{L}_{\rm GW} = -\frac{1}{2}\sum_{i=1}^{N}
\left[\frac{d_L^{\rm obs}(z_i) - d_L(z_i;\,H_0,\Omega_{m,0})}
{\sigma_{d_L,i}}\right]^2,
\label{eq:likeGW}
\end{equation}
where $d_L^{\rm obs}$ is the simulated distance, $d_L(z_i;\cdot)$ the model prediction, and $\sigma_{d_L,i}$ the total error of Eq.~\eqref{eq:total}.

\subsection{Real electromagnetic data}

We combine the LISA mock with real probes that anchor the expansion history from the recombination epoch down to low redshift.

\paragraph{Type Ia supernovae.} Two SNe~Ia compilations enter, one per scenario. Pantheon+SH0ES provides $1701$ light curves of $1550$ supernovae over $0.001 < z < 2.26$, with the SH0ES Cepheid calibration anchoring the absolute magnitude~\mbox{\cite{scolnic2022pantheon+}}. Union3 provides an independent compilation with a distinct treatment of systematics, in the form of distance moduli at $22$ spline nodes~\mbox{\cite{rubin2025union3}}; its absolute magnitude $M_B$ is marginalised analytically~\mbox{\cite{goliath2001supernovae}}. We use each in the scenario built on its preferred cosmology, and the \textit{Planck} reference scenario contains no SN~Ia data.

\paragraph{Baryon acoustic oscillations.} We include the DESI~DR2 measurements of the BAO scale~\mbox{\cite{desi2025dr2}}, comprising $13$ measurements of $D_M/r_d$, $D_H/r_d$ and $D_V/r_d$ across seven redshift bins. The sound horizon $r_d$ is marginalised analytically under a Gaussian prior $r_d = 147.09 \pm 0.26~\mathrm{Mpc}$ from \textit{Planck}~2018.

\paragraph{Cosmic chronometers.} We add $15$ cosmic-chronometer measurements of $H(z)$ from the differential ages of passively evolving galaxies~\mbox{\cite{moresco2012improved, moresco2015raising, moresco2016sixpercent}}, with the covariance matrix of~\mbox{\cite{moresco2020setting}}, which includes the systematics from stellar population synthesis modelling (see also the review~\cite{moresco2022unveiling}). These data probe the expansion rate directly and independently of any distance calibration.

\paragraph{CMB shift parameter.} Finally, the background includes the compressed CMB shift parameter $R = \sqrt{\Omega_{m,0}}\,H_0\,D_C(z_\ast)/c$, for which we adopt the \textit{Planck}~2018 TT,TE,EE+lowE value $R = 1.7502 \pm 0.0046$~\mbox{\cite{chen2019distance}}.

\subsection{Parameters and priors}
\label{sec:priors}

The model has two free cosmological parameters, sampled with flat priors, plus nuisance parameters that are marginalised analytically rather than sampled. Table~\ref{tab:priors} collects them. The flat prior on $M_B$ is improper but harmless under analytic marginalisation, because only $\chi^2$ differences enter the inference. Each SN~Ia compilation carries its own $M_B$, so a background containing one
	compilation has two marginalised nuisances in total, which is the
	$N_{\rm params}$ used in Table~\ref{tab:summary}. The two treatments of $r_d$
	listed here are the default and the CMB-free alternative; leaving $r_d$
	unconstrained entirely is the ``none'' row of Table~\ref{tab:anchors}, and it
	changes the analysis qualitatively rather than quantitatively, since the BAO then
	measure $H_0 r_d$ instead of $H_0$.

\begin{table}[t]
\centering
\caption{Every parameter of the analysis, with its prior and its role. Only the
first two are sampled. The middle block is marginalised in closed form at every
likelihood call, so those parameters never enter a chain (the Gaussian prior on
$\omega_b$ is used only in the CMB-free treatment of $r_d$ of Eq.~\eqref{eq:rdfit}).
The bottom block is held fixed.}
\label{tab:priors}
\small
\setlength{\tabcolsep}{4.5pt}
\begin{tabular}{llll}
\hline\hline
Parameter & Prior & Range / value & Applies to \\
\hline
$H_0$ [km s$^{-1}$ Mpc$^{-1}$] & flat & $[50,\,90]$ & sampled \\
$\Omega_{m,0}$ & flat & $[0.10,\,0.60]$ & sampled \\
\hline
$M_B$ & flat (marg.)~\cite{goliath2001supernovae} & improper & each SN~Ia compilation \\
$r_d$ [Mpc] & Gaussian (marg.) & $147.09 \pm 0.26$ & DESI DR2 BAO \\
$\omega_b$ & Gaussian & $0.02218 \pm 0.00055$ & DESI BAO via Eq.~\eqref{eq:rdfit} \\
\hline
$\omega_b$ (fixed) & --- & $0.02237$ & $z_\ast$~\cite{hu1996small}; value from~\cite{aghanim2020planck} \\
$N_{\rm eff}$ (fixed) & --- & $3.044$ & radiation, $r_d$ \\
$\Omega_k$ (fixed) & --- & $0$ & flat $\Lambda$CDM \\
\hline
\end{tabular}
\end{table}

Two entries need comment. The absolute magnitude $M_B$ is degenerate with $H_0$: the supernova apparent magnitude depends on the two only through the combination $M_B - 5\log_{10}H_0$, so marginalising $M_B$ with a flat prior marginalises $H_0$ along with it. This is why an SN~Ia compilation alone cannot set a fiducial $H_0$, as discussed in Section~\ref{sec:LCDM:whichfid}.

The sound horizon $r_d$ can be treated two ways, and the choice matters for how much of the result is inherited from the CMB. The default marginalises $r_d$ under the \textit{Planck} Gaussian prior quoted above. Alternatively, $r_d$ can be computed from the DESI DR2 fitting formula~\mbox{\cite{desi2025dr2, brieden2023model}},
\begin{equation}
r_d = 147.05~\mathrm{Mpc}
\left(\frac{\omega_b}{0.02236}\right)^{-0.13}
\left(\frac{\omega_{bc}}{0.1432}\right)^{-0.23}
\left(\frac{N_{\rm eff}}{3.04}\right)^{-0.1},
\label{eq:rdfit}
\end{equation}
with $\omega_{bc} = \Omega_{m,0}h^2$ and $\omega_b$ from big-bang nucleosynthesis~\mbox{\cite{schoneberg2024bbn}}. In this second form $r_d$ is not a number but a function of the sampled cosmology, and the BAO scale is calibrated without any CMB input. We quote results for the default and report the shift induced by the alternative in Section~\ref{sec:results:fiducial}.

Both points are worth making quantitative. With $M_B$ marginalised the supernova $\chi^2$ is flat in $H_0$ to within $\Delta\chi^2 \simeq 10^{-3}$ across the whole prior, a range of $40~\mathrm{km\,s^{-1}\,Mpc^{-1}}$; the residual is numerical, from the radiation term breaking the exact scaling $d_L \propto 1/H_0$. This is an exact degeneracy, not a weak constraint, and no amount of supernova data would narrow it. Only the Cepheid-calibrated Pantheon+SH0ES likelihood is peaked in $H_0$, because SH0ES supplies $M_B$ externally. As for the ruler, replacing the \textit{Planck} prior on $r_d$ by Eq.~\eqref{eq:rdfit} with a BBN prior on $\omega_b$ moves the BAO-only constraint from $H_0 = 69.04 \pm 0.51$ to $68.63 \pm 0.54$ and leaves $\Omega_{m,0}$ unchanged. It also makes $r_d$ a function of the sampled cosmology through $\omega_{bc} = \Omega_{m,0}h^2$ rather than a fixed number, which removes part of the $H_0$--$\Omega_{m,0}$ correlation that holding $r_d$ fixed induces.

For each scenario, the real-data background combines these probes, and the joint analysis multiplies their likelihood with Eq.~\eqref{eq:likeGW}. We sample the $(H_0,\Omega_{m,0})$ parameter space with the affine-invariant MCMC sampler \texttt{emcee}~\mbox{\cite{goodman2010ensemble, foreman2013emcee}}, using $64$ walkers.

Two features of this background deserve emphasis, because they largely determine the results of Section~\ref{sec:results}. The shift parameter constrains $R$ to $0.26\%$, and the BAO distances are calibrated by a \textit{Planck} prior on $r_d$ of comparable precision. Both are early-Universe anchors, and together they carry substantially more weight on $H_0$ than the SH0ES Cepheid calibration does. The background of every scenario therefore prefers a value of $H_0$ close to the \textit{Planck} one, largely irrespective of which SN~Ia compilation it contains. Section~\ref{sec:results:fiducial} quantifies how much each anchor actually contributes, and the answer is markedly asymmetric: the calibration of $r_d$ does nearly all the work, and the shift parameter almost none.

\section{Results}
\label{sec:results}

The analysis runs over a grid. Each of the three fiducials of
Table~\ref{tab:fiducials} fixes where the LISA catalogue is simulated; each of
three SN~Ia compilations (Pantheon+SH0ES, Union3 and DES-Dovekie) fixes
which supernovae enter the real-data background. Crossing them gives nine cells,
each analysed in three steps: the LISA mock alone (``LISA-only''), the real-data
combination alone (``real''), and the two together (``joint'').

Separating the two choices is what makes the comparison interpretable. When the
fiducial also selects the supernovae, only the diagonal of the grid exists, and a
small tension has two explanations that cannot be told apart: the fiducial may
agree with the data, or the scenario may simply be being compared against its own
supernovae. The separation is also cheap. Because $\mathcal{L}_{\rm bg}$ does not
depend on the fiducial, the three backgrounds are shared across the grid and the
nine cells cost fifteen chains rather than twenty-seven.
We run each with \texttt{emcee} using $64$ walkers for $2\times10^{4}$ steps,
discarding the first $5\times10^{3}$ as burn-in. Convergence is monitored through
the Gelman--Rubin statistic~\cite{gelman1992inference} and the integrated
autocorrelation time: across all fifteen chains $\hat{R} < 1.002$ and the
effective sample size exceeds $2.8\times10^{4}$ in both parameters.
Table~\ref{tab:summary} collects the marginalised constraints,
Table~\ref{tab:tension} the internal tensions, and
Figs.~\ref{fig:planck}--\ref{fig:pantheon} the contours.

\subsection{Recovery of the fiducial}
\label{sec:results:recovery}

The first question is whether the pipeline is unbiased. Since the LISA-only posterior depends only on the fiducial, three chains answer it. For the \textit{Planck} baseline it gives $H_0 = 67.61 \pm 1.01$ and $\Omega_{m,0} = 0.3106 \pm 0.0205$, against the input $(67.34, 0.3153)$. For Union3, $H_0 = 68.14 \pm 1.01$ and $\Omega_{m,0} = 0.2944 \pm 0.0195$, against $(67.10, 0.3190)$. For Pantheon+SH0ES, $H_0 = 70.49 \pm 1.11$ and $\Omega_{m,0} = 0.3868 \pm 0.0250$, against $(73.04, 0.3340)$. All are the $68\%$ credible intervals of Table~\ref{tab:summary}. The reduced $\chi^2$ stays close to unity throughout (Table~\ref{tab:summary}).

These offsets have to be read in two dimensions rather than one. The LISA-only posterior is almost perfectly degenerate in the $(H_0,\Omega_{m,0})$ plane (the correlation coefficient is $-0.985$ in all three scenarios), so a displacement along the degeneracy costs little likelihood while looking sizeable in either marginal, and a displacement across it costs a great deal while looking modest. Evaluated with the full posterior covariance, the recovered position sits $0.07\sigma$ from the injected fiducial for \textit{Planck}18, $1.21\sigma$ for Union3 and $2.07\sigma$ for Pantheon+SH0ES. The first is far smaller than either marginal would suggest; the last is a genuine fluctuation, of the size the coverage test below leads one to expect in one scenario out of three.

\begin{figure}[!htbp]
\centering
\includegraphics[width=\linewidth]{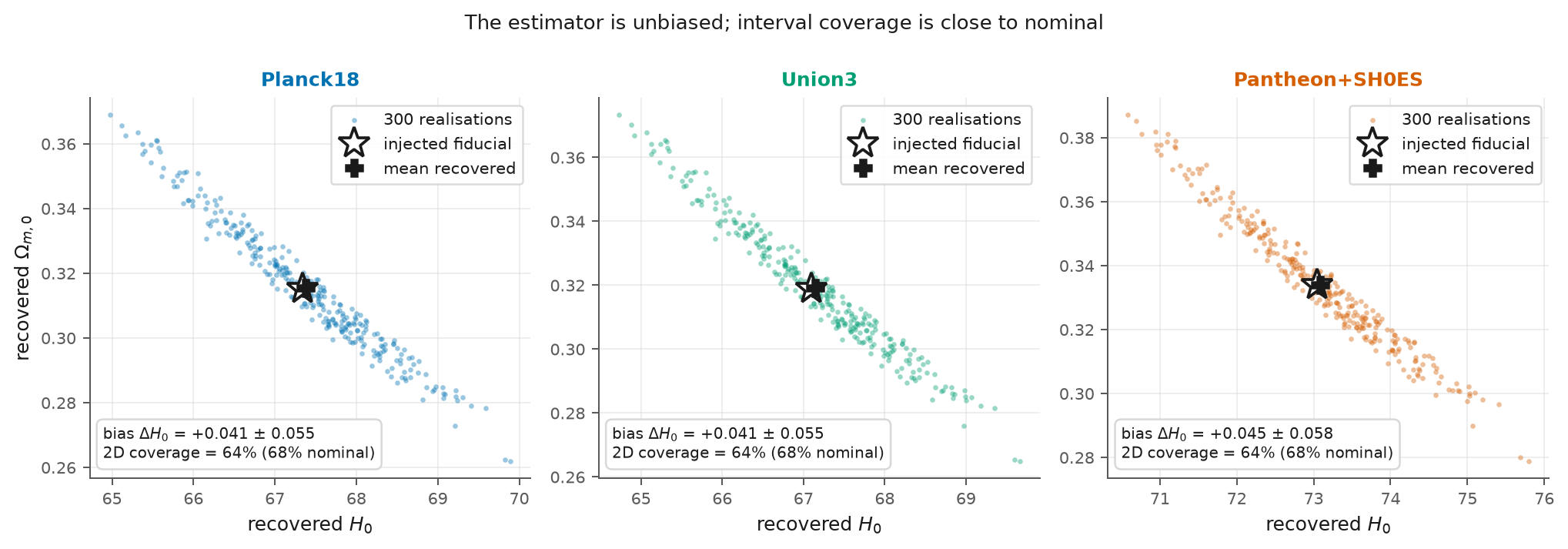}
\caption{Recovery of the injected fiducial over 300 independent noise realisations per scenario. Points scatter along the $H_0$--$\Omega_{m,0}$ degeneracy, which is why a deviation that looks large in the one-dimensional marginals can be unremarkable in two dimensions. The mean recovered value (cross) sits on the injected fiducial (star) in all three cases.}
\label{fig:bias}
\end{figure}

A single realisation, however, cannot establish that the estimator is unbiased: with correctly sized intervals about a third of realisations fall outside $1\sigma$ by construction. We therefore repeated the exercise over 300 independent noise realisations per scenario (Figure~\ref{fig:bias}). The mean recovered $H_0$ differs from the injected value by $+0.04 \pm 0.06$, consistent with zero in every scenario, and the fraction of realisations whose 68\% two-dimensional credible region contains the fiducial is $64\%$ against the nominal $68\%$. The estimator is therefore unbiased and its intervals close to correctly calibrated, which is what a forecast needs, and a stronger statement than the agreement of any single draw.

\begin{table*}[t]
\centering
\caption{Marginalised constraints from every chain of the grid. The three
\emph{Real} rows are the background likelihoods, one per SN~Ia compilation, each
combined with DESI-DR2, cosmic chronometers and Planck-18; they carry no
dependence on the fiducial and are therefore shared across the grid. The three
\emph{LISA} rows are the mock catalogues, one per fiducial; they carry no
dependence on the supernova compilation. The nine \emph{Joint} rows are the
products, one per cell. Central values and errors are the median and the $68\%$
credible interval of the marginal posterior. $\chi^2_{\rm min} = -2\max\ln\mathcal{L}$
over the chain; $\chi^2_{\rm red} = \chi^2_{\rm min}/(N_{\rm data}-N_{\rm params})$,
with $N_{\rm params}$ counting the two sampled parameters plus the analytically
marginalised nuisances of the datasets in that row.
Several joint rows fall
outside both of the posteriors they combine; Section~\ref{sec:results:tiebreak}
explains why this is expected rather than an error.}
\label{tab:summary}
\small
\setlength{\tabcolsep}{3pt}
\begin{tabular}{lcccc}
\hline\hline
Dataset & $H_0$ [km/s/Mpc] & $\Omega_{\rm m}$ & $\chi^2_{\rm min}$ & $\chi^2_{\rm red}$ \\
\hline
Real (Pantheon+) & $69.02^{+0.35}_{-0.35}$ & $0.3026^{+0.0054}_{-0.0053}$ & $1499.62$ & $0.892$ \\
Real (Union3) & $68.36^{+0.38}_{-0.38}$ & $0.3102^{+0.0059}_{-0.0059}$ & $46.19$ & $0.983$ \\
Real (DES) & $68.34^{+0.37}_{-0.36}$ & $0.3106^{+0.0056}_{-0.0056}$ & $1652.43$ & $0.896$ \\
LISA @ Planck18 & $67.61^{+1.01}_{-1.01}$ & $0.3106^{+0.0205}_{-0.0205}$ & $796.28$ & $1.040$ \\
LISA @ Union3 & $68.14^{+1.01}_{-1.00}$ & $0.2944^{+0.0194}_{-0.0195}$ & $790.21$ & $1.032$ \\
LISA @ Pantheon+ & $70.49^{+1.11}_{-1.10}$ & $0.3868^{+0.0250}_{-0.0251}$ & $803.62$ & $1.049$ \\
Joint: Planck18 + Pantheon+ & $68.34^{+0.29}_{-0.29}$ & $0.3031^{+0.0052}_{-0.0052}$ & $2310.14$ & $0.943$ \\
Joint: Planck18 + Union3 & $67.84^{+0.31}_{-0.31}$ & $0.3109^{+0.0056}_{-0.0056}$ & $849.78$ & $1.043$ \\
Joint: Planck18 + DES & $67.83^{+0.30}_{-0.30}$ & $0.3112^{+0.0054}_{-0.0054}$ & $2456.01$ & $0.940$ \\
Joint: Union3 + Pantheon+ & $68.22^{+0.29}_{-0.29}$ & $0.3021^{+0.0051}_{-0.0052}$ & $2312.86$ & $0.944$ \\
Joint: Union3 + Union3 & $67.72^{+0.31}_{-0.31}$ & $0.3099^{+0.0056}_{-0.0056}$ & $850.85$ & $1.044$ \\
Joint: Union3 + DES & $67.70^{+0.30}_{-0.30}$ & $0.3103^{+0.0054}_{-0.0054}$ & $2457.09$ & $0.940$ \\
Joint: Pantheon+ + Pantheon+ & $72.33^{+0.31}_{-0.31}$ & $0.3076^{+0.0053}_{-0.0053}$ & $2685.63$ & $1.096$ \\
Joint: Pantheon+ + Union3 & $72.19^{+0.33}_{-0.33}$ & $0.3097^{+0.0057}_{-0.0057}$ & $1256.61$ & $1.542$ \\
Joint: Pantheon+ + DES & $72.17^{+0.32}_{-0.32}$ & $0.3101^{+0.0055}_{-0.0055}$ & $2862.87$ & $1.096$ \\
\hline
\end{tabular}
\end{table*}

\subsection{LISA as a tiebreaker}
\label{sec:results:tiebreak}

\begin{figure}[!htbp]
\centering
\includegraphics[width=0.86\linewidth]{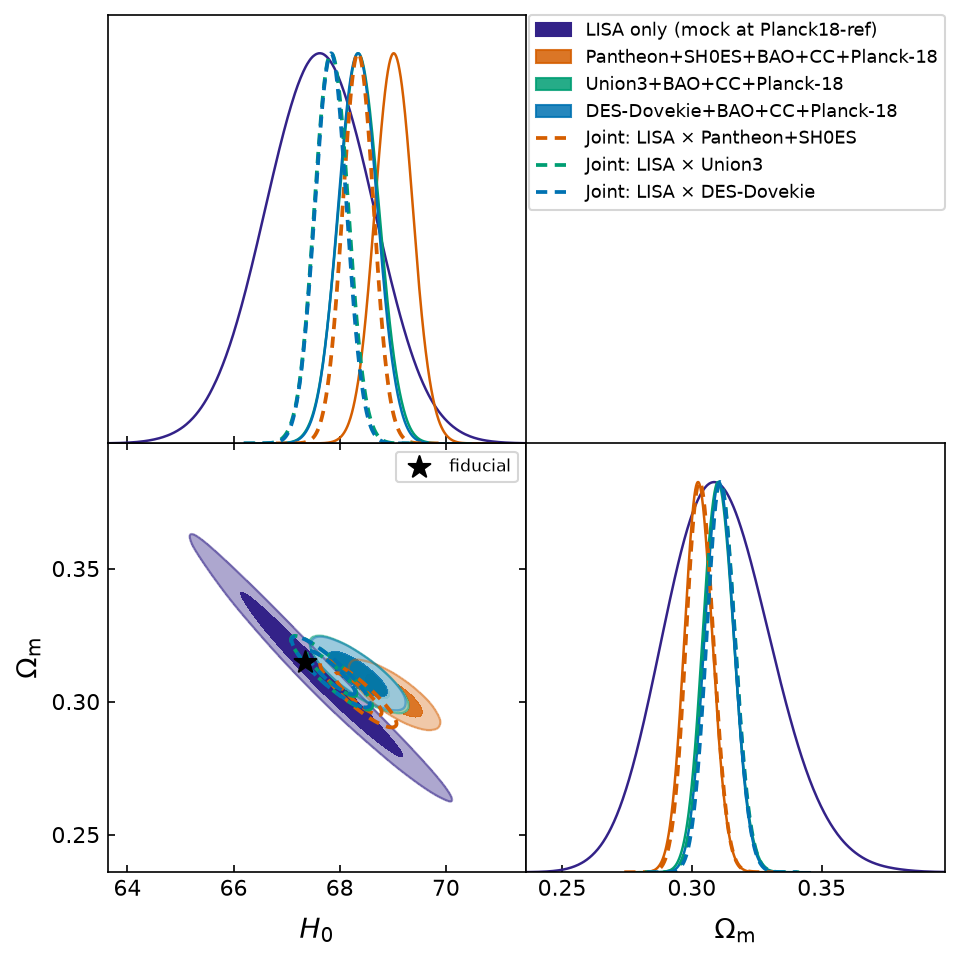}
\caption{\textit{Planck}18 fiducial. Marginalised $(H_0,\Omega_{m,0})$
posteriors, with $68\%$ and $95\%$ contours. \emph{Filled}: the LISA-only mock
(purple) and the three real-data backgrounds, each being one SN~Ia compilation
combined with DESI-DR2, cosmic chronometers and Planck-18: Pantheon+SH0ES
(orange), Union3 (green), DES-Dovekie (blue). \emph{Dashed}: the joint
posterior of the mock with each background, drawn in the colour of the
compilation that entered it. The star marks the fiducial at which the mock was
generated. The offset between the purple contour and each coloured one is the
internal tension of Table~\ref{tab:tension}. The Union3 and DES-Dovekie curves
coincide to within the line width (the green is largely hidden beneath the
blue) because with $M_B$ marginalised neither compilation carries
information on $H_0$, so both inherit it from the DESI and Planck-18
components they share.}
\label{fig:planck}
\end{figure}

\begin{figure}[!htbp]
\centering
\includegraphics[width=0.86\linewidth]{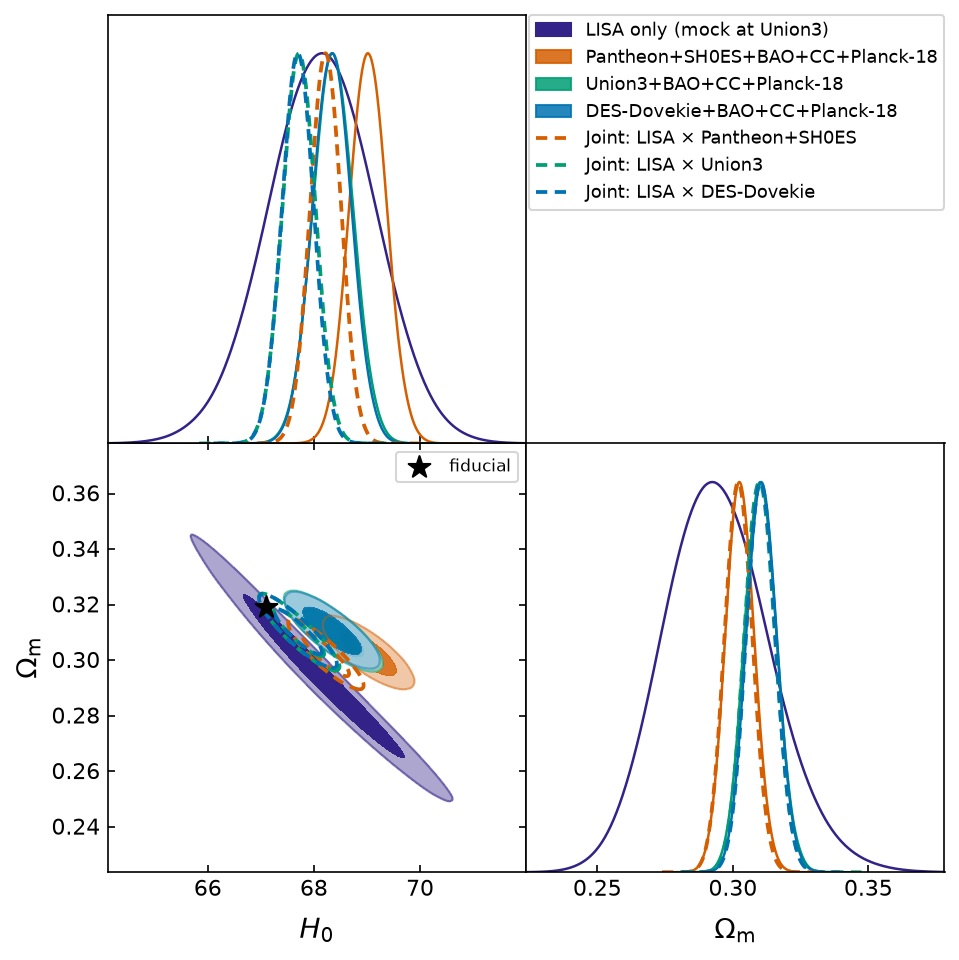}
\caption{Union3 fiducial; colours and line styles as in
Figure~\ref{fig:planck}. The three backgrounds are identical to those of
Figure~\ref{fig:planck}, because they do not depend on the fiducial, and the only
things that move between the two figures are the purple contour and the three
joints that follow it. That the displacement is small is the point: the two
fiducials are separated by $0.24~\mathrm{km\,s^{-1}\,Mpc^{-1}}$ in $H_0$, a
quarter of the LISA-only uncertainty.}
\label{fig:union3}
\end{figure}

\begin{figure}[!htbp]
\centering
\includegraphics[width=0.86\linewidth]{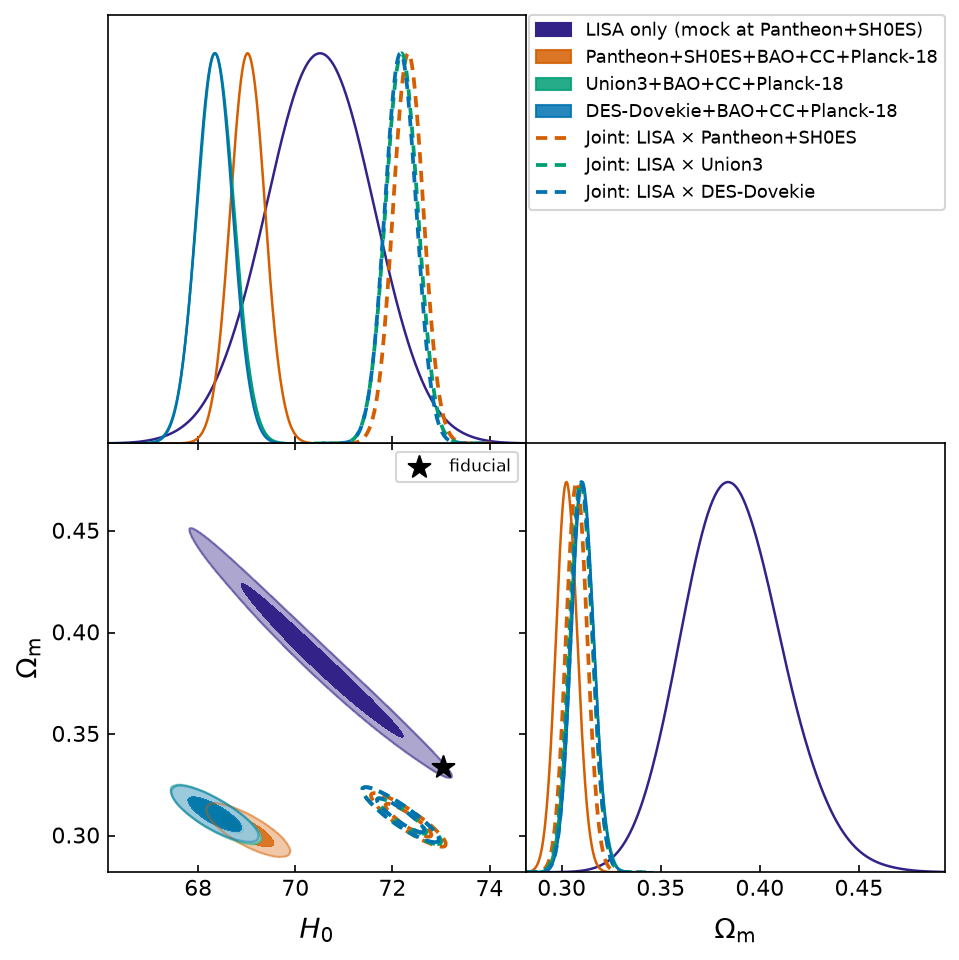}
\caption{Pantheon+SH0ES fiducial; colours and line styles as in
Figure~\ref{fig:planck}. Here the LISA-only contour is disjoint from all three
backgrounds, and the separation runs across the $H_0$--$\Omega_{m,0}$
degeneracy rather than along it, which is why the two-dimensional tension of
Table~\ref{tab:tension} is so much larger than either marginal. Two features
are discussed in the text: the fiducial (star) sits at the edge of the LISA
contour, the $2.1\sigma$ recovery of Section~\ref{sec:results:recovery}; and
the three joint posteriors (dashed) fall outside \emph{both} of the
distributions they combine, at $H_0 \simeq 72$, which is the expected
behaviour of a product of misaligned anisotropic likelihoods and not a
sampling artefact (Section~\ref{sec:results:tiebreak}).}
\label{fig:pantheon}
\end{figure}

The central result is the consistency between the simulated LISA data and the
real measurements, quantified by the two-dimensional tension $n_\sigma$ in the
$(H_0,\Omega_{m,0})$ plane between the LISA-only and real-data posteriors.
Table~\ref{tab:tension} gives all nine values; Figs.~\ref{fig:planck}--\ref{fig:pantheon}
show, for each fiducial, the LISA-only contour against all three backgrounds at
once. The grid separates two effects that a diagonal-only design conflates, and
they differ by more than an order of magnitude.

\emph{The fiducial dominates.} Across the grid the tension runs from
$2.08\sigma$ up to values far beyond the regime in which a Gaussian $\sigma$ is a meaningful number, and essentially the whole of that range is set by
where the catalogue was simulated. The \textit{Planck} fiducial gives
$2.08$--$3.22\sigma$, Union3 gives $3.25$--$4.24\sigma$, and Pantheon+SH0ES
gives values well in excess of $5\sigma$ throughout. The last is not a marginal preference: a catalogue
simulated in a Cepheid-calibrated universe is irreconcilable with every
background we tested. The one-dimensional marginals understate it badly (they
give $1.26$--$1.84\sigma$ in $H_0$ and $2.94$--$3.26\sigma$ in $\Omega_{m,0}$),
because both posteriors are elongated along the $H_0$--$\Omega_{m,0}$ degeneracy
while the offset between them runs across it (Fig.~\ref{fig:pantheon}). This is
the expected consequence of a fiducial sitting some $4~\mathrm{km\,s^{-1}\,Mpc^{-1}}$
above the real-data combination in $H_0$ while the LISA catalogue faithfully
reproduces the cosmology it was built on.

A tension of this size could also point to a numerical problem, so we
recomputed it without sampling and without noise. For each fiducial we build
the Fisher matrix of the catalogue, $F = J^{\rm T}\,{\rm diag}(\sigma_{d_L,i}^{-2})\,J$
with $J_{i\alpha} = \partial d_L(z_i)/\partial\theta_\alpha$ evaluated at the
fiducial, take the background covariance from the Hessian of its $\chi^2$ at the
best fit, and evaluate $Q_{\rm DM}$ with the fiducial in place of the LISA
posterior mean. For the Pantheon+SH0ES
fiducial against the Pantheon+SH0ES and Union3 backgrounds this again gives values far in excess of $5\sigma$, and between $2.3$
and $3.8$ for the other two fiducials, in agreement with
Table~\ref{tab:tension} within the scatter between noise realisations. The origin of the large number is simple. Weighted by
their errors, the events constrain the distance around $z \simeq 1.5$, and the
stacked catalogue measures $d_L$ there to about $0.25\%$, as expected from a
median single-event error of $8\%$ and $768$ events. Raising $H_0$ and raising
$\Omega_{m,0}$ both shorten $d_L$ at these redshifts, and the Pantheon+SH0ES
fiducial lies above the background in both parameters, so its distances are
shorter than those preferred by the background by about $8\%$ (median over the
catalogue). In the eigenbasis of the LISA covariance this offset is large
across the degeneracy direction and modest along it, and adding the
background covariance drives $Q_{\rm DM}$ well beyond the range in which its
conversion to a Gaussian $n_\sigma$ remains meaningful. The \textit{Planck} and Union3 fiducials, in contrast, differ from the
background distances by only $1.1$ to $1.6\%$. Finally, the conversion of
$Q_{\rm DM}$ into $n_\sigma$ relies on the Gaussian tail of the $\chi^2$
distribution, which cannot be trusted much beyond $5\sigma$. Values
above that point are reported here only as ``beyond $5\sigma$'' and
should be read as strong incompatibility rather than as precise
numbers; the quantity that carries the result is $Q_{\rm DM}$ itself.

\emph{The supernova compilation barely matters.} Within a fiducial, changing the
compilation moves the tension by at most $1.2\sigma$: $2.08$, $2.09$ and
$3.22\sigma$ for Union3, DES-Dovekie and Pantheon+SH0ES at the \textit{Planck}
fiducial, and $3.25$, $3.28$ and $4.24\sigma$ at the Union3 fiducial. Union3 and
DES-Dovekie are, for this purpose, the same dataset: their tensions differ by
$0.01$, $0.03$ and $0.10\sigma$ in the three fiducials, and their backgrounds
return $H_0 = 68.36 \pm 0.38$ and $68.34 \pm 0.37$ with $\Omega_{m,0} = 0.3102
\pm 0.0059$ and $0.3106 \pm 0.0056$. The agreement is structural, not
coincidental: both marginalise $M_B$, so neither contributes any information on
$H_0$ (Section~\ref{sec:priors}), and the $H_0$ of their backgrounds comes
entirely from DESI and Planck-18, which they share. Only Pantheon+SH0ES stands
apart, and only because the Cepheid calibration gives it an independent handle on
$H_0$, pulling its background up to $69.02 \pm 0.35$.

The diagonal of the grid behaves as one would expect, and the size of the effect
is worth recording because a diagonal-only design cannot measure it. For both
fiducials that have an ``own'' compilation, that cell is the minimum of its row
($3.25\sigma$ for Union3 and, for Pantheon+SH0ES, a value well beyond $5\sigma$), but the
margin over the next-best compilation is $0.03\sigma$ and $0.85\sigma$
respectively, a small fraction of the difference between fiducials. Pairing each
scenario with its own supernovae flatters it, by an amount too small to matter.

\emph{The two low-tension fiducials are not separated.} Against a common
background, the \textit{Planck} fiducial gives $2.08\sigma$ and Union3
$3.25\sigma$, and the difference carries little weight: independent noise realisations move each
number by $\pm 0.4\sigma$ (Section~\ref{sec:results:fiducial}), and the two
fiducials themselves differ by only $0.24$ in $H_0$ and $0.004$ in
$\Omega_{m,0}$, about a quarter and a fifth, respectively, of the LISA-only
uncertainties. A measurement cannot rank two hypotheses separated
by a small fraction of its own resolution.

One feature of Table~\ref{tab:summary} may look like an
error but is not. In five of the nine cells the joint posterior falls outside
\emph{both} of the posteriors it combines; at the Pantheon+SH0ES fiducial the
joint gives $H_0 = 72.33 \pm 0.31$, above both the LISA-only value of $70.49$ and
the background value of $69.02$. This is the expected behaviour of a product of
two strongly anisotropic two-dimensional likelihoods whose degeneracy directions
are misaligned: the combined mean is a weighting by precision \emph{matrices}
rather than by componentwise variances, and such a weighting need not place each
component between the corresponding components of the two inputs. We verified that the sampled joints
reproduce the analytic Gaussian product in all nine cells to better than
$0.3~\mathrm{km\,s^{-1}\,Mpc^{-1}}$. The effect is a symptom of the tension, not
of the sampler, and it is largest exactly where the two inputs disagree most.

This result is not an artefact of the catalogue size we adopt. More recent population models predict far fewer events with an identified electromagnetic counterpart than the $768$ assumed here, of order $2$ to $20$ in four years~\cite{mangiagli2022massive, mangiagli2024massive}; for $N = 10$ and $20$ the Fisher estimate described above gives $3.3\sigma$ and $4.9\sigma$ for the Pantheon+SH0ES fiducial against the Union3 background, so the exclusion is not an artefact of assuming a large catalogue.

The tiebreaker reading must therefore be stated with care. A LISA bright-siren catalogue of this quality would rule out a universe whose expansion history matches the Cepheid-calibrated Pantheon+SH0ES cosmology, by a wide margin and with far fewer events than we simulate here. It would not, on this error budget, decide between the \textit{Planck} and Union3 scenarios. This is not because the measurement is weak, but because those two cosmologies lie too close together for any standard-siren catalogue of plausible size to tell apart. The first statement is the informative one; presenting the three tensions as a ranking would over-read the second.

\begin{table*}[t]
\centering
\caption{Internal tension between the LISA-only posterior and each real-data
background, over the full grid of three fiducials by three SN~Ia compilations.
$Q_{\rm DM}$ is the Raveri--Hu difference-in-mean
statistic~\cite{raveri2019concordance}, built from the difference between the two
posterior means and the sum of their covariances; $n_\sigma$ (2D) converts it
through the $\chi^2$ distribution with two degrees of freedom. The
one-dimensional entries are given for reference and understate the discrepancy
badly, because both posteriors are elongated along the $H_0$--$\Omega_{m,0}$
degeneracy while the offset between them runs across it; they should not be read
as the result. Rows are grouped by fiducial. Reading down a group asks which
compilation a catalogue simulated at that fiducial would agree with; reading
across groups at fixed compilation isolates the effect of the fiducial. The
second comparison is the informative one: the fiducial is what drives the
tension, the compilation moves it by at most $1.2\sigma$. Where the
two-dimensional $n_\sigma$ would formally exceed $\sim 5$, we report it only
as ``$\gg 5$'', since the Gaussian-tail conversion from $Q_{\rm DM}$ is not
meaningful that far out and the value should be read as strong
incompatibility rather than as a precise number (see
Section~\ref{sec:results:tiebreak}).}
\label{tab:tension}
\small
\setlength{\tabcolsep}{4.5pt}
\begin{tabular}{llccccc}
\hline\hline
LISA fiducial & $(H_0, \Omega_{\rm m})$ & SNe in background & $Q_{\rm DM}$ & $n_\sigma$ (2D) & $n_\sigma$ ($H_0$) & $n_\sigma$ ($\Omega_{\rm m}$) \\
\hline
Planck18-ref & $(67.34, 0.3153)$ & Pantheon+SH0ES & $13.29$ & $3.22$ & $1.31$ & $0.38$ \\
Planck18-ref & $(67.34, 0.3153)$ & Union3 & $6.58$ & $2.08$ & $0.69$ & $0.02$ \\
Planck18-ref & $(67.34, 0.3153)$ & DES-Dovekie & $6.64$ & $2.09$ & $0.67$ & $0.00$ \\
Union3 & $(67.10, 0.3190)$ & Pantheon+SH0ES & $21.45$ & $4.24$ & $0.83$ & $0.40$ \\
Union3 & $(67.10, 0.3190)$ & Union3 & $13.56$ & $3.25$ & $0.20$ & $0.77$ \\
Union3 & $(67.10, 0.3190)$ & DES-Dovekie & $13.72$ & $3.28$ & $0.19$ & $0.79$ \\
Pantheon+SH0ES & $(73.04, 0.3340)$ & Pantheon+SH0ES & $334.23$ & $\gg 5$ & $1.26$ & $3.26$ \\
Pantheon+SH0ES & $(73.04, 0.3340)$ & Union3 & $365.77$ & $\gg 5$ & $1.82$ & $2.94$ \\
Pantheon+SH0ES & $(73.04, 0.3340)$ & DES-Dovekie & $369.54$ & $\gg 5$ & $1.84$ & $2.94$ \\
\hline
\end{tabular}
\end{table*}

\subsection{Joint constraints}

Combining the LISA mock with the real data tightens the constraints, but only where the combination is legitimate. At the \textit{Planck}18 fiducial the joint fit against a Union3 background gives $H_0 = 67.84 \pm 0.31$ and $\Omega_{m,0} = 0.3109 \pm 0.0056$, improving on the background alone ($\pm 0.38$) by about $18\%$ and on LISA alone ($\pm 1.0$) by a factor of three; the DES-Dovekie background gives $67.83 \pm 0.30$ and $0.3112 \pm 0.0054$, indistinguishable from it, as the near-identity of the two compilations implies. The Union3 fiducial behaves the same way, at $67.72 \pm 0.31$. Reduced $\chi^2$ stays between $0.94$ and $1.04$ across these six cells. The gain is modest because the background already contains Planck-18 and a \textit{Planck}-calibrated BAO ruler, which anchor $H_0$ tightly; what LISA adds is an independent, calibration-free distance scale rather than a large statistical improvement.

For the Pantheon+SH0ES fiducial we report the joint fits for completeness but do not interpret them. Combining datasets that disagree at a significance far beyond $5\sigma$ yields a compromise, $H_0 \simeq 72.2$--$72.3$, more than $3~\mathrm{km\,s^{-1}\,Mpc^{-1}}$ from every background, that describes neither, and the reduced $\chi^2$ rises to $1.10$, or $1.54$ against the Union3 background, from $0.89$--$0.98$ for the backgrounds alone. A joint constraint is meaningful only once the datasets entering it have been shown to be consistent, which is precisely what the previous subsection tests.

\subsection{How much depends on the fiducial?}
\label{sec:results:fiducial}

Because the mock is generated at the scenario's fiducial and the LISA-only posterior recovers it, the internal tension is close to a statement about where the fiducial sits relative to the electromagnetic data (Section~\ref{sec:LCDM:howused}). It is therefore worth asking how much of the result is a property of that choice. We repeat the measurement holding the real-data background fixed at a common reference (BAO + cosmic chronometers + Planck-18) and varying only the fiducial at which the LISA catalogue is generated.

The fiducial choice spans the whole range of possible conclusions: the internal tension runs from $0.01\sigma$ up to values far beyond $5\sigma$ across the fiducials tested, with identical simulated data in every case. Fiducials drawn from multi-probe combinations cluster tightly, at $H_0 \simeq 68.3$--$69.6$, and all yield tensions below $1.1\sigma$; this is expected rather than reassuring, since those combinations are dominated by the same BAO and CMB information that defines the reference, so agreement is partly built in. Only the Cepheid-calibrated Pantheon+SH0ES fiducial sits far enough away to produce a large tension. Repeating the noise realisation shows a scatter of $\pm 0.4\sigma$ on any single number, so differences below about $1\sigma$ between scenarios are not resolved by one catalogue.

Adopting Eq.~\eqref{eq:rdfit} instead of the \textit{Planck} prior on $r_d$ moves the BAO-calibrated fiducial from $H_0 = 69.04 \pm 0.51$ to $68.63 \pm 0.54$, with $\Omega_{m,0}$ unchanged. The shift is modest, but it removes the CMB from the calibration chain, and we quote it because the size of the internal tension in the low-tension scenarios is comparable to it.

Table~\ref{tab:anchors} pushes this further by rebuilding the real-data background without each primordial anchor in turn, holding the three LISA-only posteriors frozen so that every difference across a row comes from the background alone. Two things stand out, and they are markedly asymmetric. Removing Planck-18 changes every tension by less than $0.4\sigma$ and the background $H_0$ by $0.3$: once DESI BAO enter with a calibrated ruler, the compressed CMB datum is very nearly redundant. Replacing the \textit{Planck} prior on $r_d$ by Eq.~\eqref{eq:rdfit} with a BBN prior on $\omega_b$, which removes CMB information from the analysis altogether, lowers the tensions by a factor of about two, yet still leaves Pantheon+SH0ES well above $5\sigma$. The exclusion therefore does not rest on the CMB.

Removing the absolute calibration of the BAO ruler is a different matter, and it delimits the claim usefully. With a flat prior on $r_d$ the BAO constrain the product $H_0 r_d$ rather than $H_0$, the background uncertainty grows from $\sigma(H_0) = 0.37$ to $3.87$, and the two low-tension fiducials collapse below $0.5\sigma$ simply because the reference no longer measures anything. The Pantheon+SH0ES case is the informative one: with no ruler calibration the Cepheid ladder becomes the only absolute distance scale available, the background built on it moves to $H_0 = 73.41 \pm 0.98$, and the tension falls to $2.8\sigma$. What the tiebreaker measures, then, is the consistency of a LISA catalogue with a background whose distance scale is set by early-Universe physics. That this comparison is decisive is a result; that it is a comparison against an early-Universe-anchored reference is a premise, and worth stating as one.

\begin{table}[t]
\centering
\caption{How much of the tension is inherited from the early Universe. The
real-data background is rebuilt in five configurations of its two primordial
anchors while the three LISA-only posteriors are held \emph{frozen}, so that
every difference across a row comes from the background alone. ``Planck-18''
denotes the compressed CMB shift parameter $R$ of Section~\ref{sec:data}; for $r_d$,
\textit{Planck} is the Gaussian prior $147.09 \pm 0.26~\mathrm{Mpc}$, ``BBN''
replaces it by the fitting formula of Eq.~\eqref{eq:rdfit} with $\omega_b$ from
big-bang nucleosynthesis, and ``none'' leaves $r_d$ free, in which case the BAO
constrain $H_0 r_d$ rather than $H_0$. The supernova compilation is held fixed at
Union3, for the reasons given in Section~\ref{sec:results:tiebreak}; the last two columns give the background
$H_0$ under Union3 and, for contrast, under Pantheon+SH0ES, which is the only
compilation with an absolute calibration of its own. Chains are shorter than in
production, so the first row reproduces Table~\ref{tab:tension} to within the
chain noise. Entries beyond $\sim 5\sigma$ are reported only as ``$>5$'', since
the Gaussian-tail conversion is not meaningful that far out
(Section~\ref{sec:results:tiebreak}). Removing Planck-18 changes nothing; removing the calibration of the
BAO ruler changes everything.}
\label{tab:anchors}
\small
\setlength{\tabcolsep}{4pt}
\begin{tabular}{llcccc}
\hline\hline
Planck-18 & $r_d$ & Planck18-ref & Union3 & Pantheon+SH0ES & $H_0^{\rm bg}$ (Union3) \\
\hline
yes & \textit{Planck} & $2.09$ & $3.24$ & $>5$ & $68.36 \pm 0.37$ \\
no  & \textit{Planck} & $2.17$ & $3.23$ & $>5$ & $68.68 \pm 0.49$ \\
yes & BBN             & $1.11$ & $1.63$ & $>5$  & $68.63 \pm 0.54$ \\
yes & none            & $0.05$ & $0.40$ & $2.81$  & $68.77 \pm 3.87$ \\
no  & none            & $0.07$ & $0.14$ & $2.96$  & $68.46 \pm 3.94$ \\
\hline
\end{tabular}
\end{table}

\begin{figure}[!htbp]
\centering
\includegraphics[width=\linewidth]{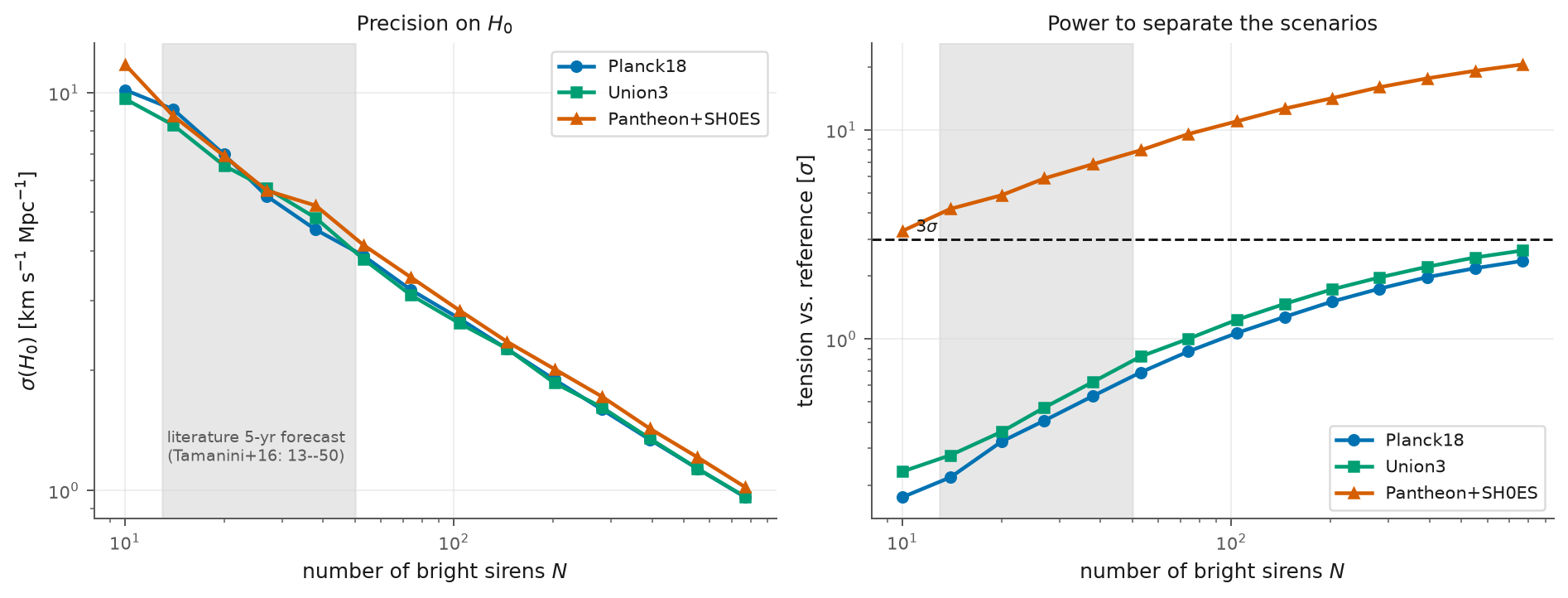}
\caption{Discriminating power against catalogue size. \emph{Left}: precision on $H_0$; each point averages 25 random subsamples of the catalogue. \emph{Right}: tension between the LISA-only posterior at each fiducial and the fixed real-data reference. The shaded band marks the range of bright sirens forecast for a five-year mission in~\mbox{\cite{tamanini2016science}}; more recent models predict fewer~\cite{mangiagli2022massive}.}
\label{fig:power}
\end{figure}

The complementary question is how much of this depends on the size of the catalogue. Figure~\ref{fig:power} recasts the forecast as a function of the number of bright sirens. The precision on $H_0$ follows the expected $N^{-1/2}$, reaching $\sigma(H_0) \simeq 1$ at $N = 768$ but only $\simeq 4$--$6$ over the range of early forecasts for a five-year mission. In that range the Pantheon+SH0ES fiducial is already separated from the reference at more than $3\sigma$, while the Planck18 and Union3 fiducials stay below $3\sigma$ even at $N = 768$. Resolving a $5.7~\mathrm{km\,s^{-1}\,Mpc^{-1}}$ offset at $3\sigma$ needs of order 200 bright sirens on this error budget. That number, rather than the ranking of three fiducials, is the forecast we regard as robust.

\begin{figure}[!htbp]
\centering
\includegraphics[width=\linewidth]{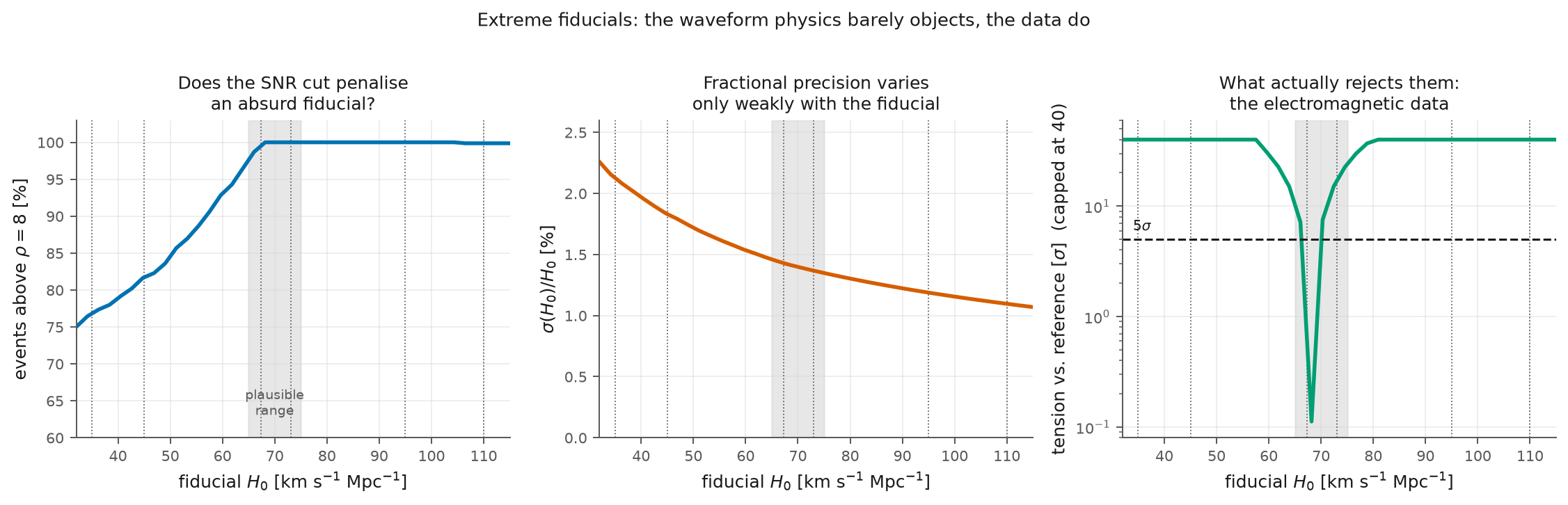}
\caption{Behaviour at deliberately unphysical fiducials, with $\Omega_{m,0}$ held at the \textit{Planck} value and only $H_0$ varied. The shaded band marks the range spanned by present measurements; dotted lines mark the fiducials tabulated in the text. \emph{Left}: fraction of the catalogue still above the detection threshold $\rho = 8$. Lowering $H_0$ lengthens every distance, lowers the signal-to-noise ratio and removes events; raising it costs nothing. \emph{Centre}: fractional precision on $H_0$, which varies only weakly, so the mock recovers whatever fiducial it was built on. \emph{Right}: internal tension against the fixed real-data reference; values far beyond $5\sigma$ are capped for display, since a Gaussian $\sigma$ is not a meaningful measure that far into the tail. The rejection of extreme fiducials comes from the electromagnetic data, not from the gravitational-wave physics.}
\label{fig:extreme}
\end{figure}

A sharper version of the same question is whether the gravitational-wave side constrains the fiducial at all. We therefore repeated the construction at deliberately unphysical fiducials, from $H_0 = 35$ to $115~\mathrm{km\,s^{-1}\,Mpc^{-1}}$ (Figure~\ref{fig:extreme}). There is a real effect, and it is one-sided. Because the signal-to-noise ratio scales as $\rho \propto 1/d_L$ at fixed source parameters, lowering $H_0$ stretches every distance and pushes events below the detection threshold: at $H_0 = 45$ only $79\%$ of the catalogue still satisfies $\rho > 8$, and at $H_0 = 35$ only $74\%$, with the median fractional distance error rising from $7.9\%$ to $12.7\%$. Raising $H_0$ costs nothing, since shorter distances only improve the signal-to-noise ratio.

The effect is, however, far too weak to act as a physical prior. The fractional precision on the Hubble constant, $\sigma(H_0)/H_0$, stays between $1.1\%$ and $2.3\%$ across a range in which $H_0$ itself varies by a factor of three, so a catalogue generated at $H_0 = 35$ recovers $35$ just as cleanly as one generated at $67$ recovers $67$. The standard-siren measurement is calibration-free, but it is not self-validating. Nothing in the waveform knows that a given expansion history is implausible. What rejects the extreme fiducials is the comparison with the electromagnetic data (the internal tension becomes extremely large at both ends, far beyond the regime in which a Gaussian $\sigma$ is meaningful), and that is a statement about the real datasets, not about LISA. It reinforces the reading of Section~\ref{sec:LCDM:howused}: the simulated data supply the error budget, and the discriminating power that follows from it, but the fiducial is an assumption of the exercise throughout.

\section{Conclusions}
\label{sec:conclusions}

We have forecast the cosmological reach of LISA bright sirens, treating the mission not as a stand-alone ruler but as a tiebreaker among cosmological scenarios that are currently in tension. We built mock catalogues of massive black hole binary bright sirens under three fiducial cosmologies (the \textit{Planck}~2018 baseline, Union3, and Pantheon+SH0ES), each catalogue carrying a self-consistent error budget from instrumental noise, weak lensing, and peculiar velocities. We then compared the simulated data with a real combination of SNe~Ia, DESI baryon acoustic oscillations, cosmic chronometers and the compressed \textit{Planck}~2018 CMB, crossing each fiducial with each of three supernova compilations so that the two choices could be told apart.

Three results stand out. First, the pipeline is unbiased: over $300$ independent noise realisations per scenario the mean recovered $H_0$ differs from the injected value by $+0.04 \pm 0.06$, and $64\%$ of realisations contain the fiducial within their $68\%$ two-dimensional credible region, against the nominal $68\%$.

Second, the fiducial and not the supernova compilation sets the answer. Across the nine combinations the internal tension runs from $2.08\sigma$ up to values well beyond $5\sigma$; the fiducial is what drives essentially all of that range, the compilation moves it by at most $1.2\sigma$. Union3 and DES-Dovekie are statistically indistinguishable throughout, their tensions differing by at most $0.10\sigma$, because with $M_B$ marginalised neither carries information on $H_0$ and both inherit it from the same BAO and CMB components; only Pantheon+SH0ES separates, and only through its Cepheid calibration. Pairing each fiducial with its own supernovae flatters it, but by $0.03\sigma$ and $0.85\sigma$, an effect that a diagonal-only design cannot measure and that turns out to be negligible.

Third, the discriminating power is strongly asymmetric. A LISA catalogue generated in a Pantheon+SH0ES universe is inconsistent with every background we test, at a significance far beyond $5\sigma$, and the conclusion survives a large reduction in catalogue size: it still exceeds $5\sigma$ with $50$ bright sirens, and a Fisher estimate gives $3.3$ to $4.9\sigma$ for the $10$ to $20$ events of recent population models. The \textit{Planck}~2018 and Union3 fiducials, at $2.1\sigma$ and $3.3\sigma$ against a common background, are not separated by the measurement and cannot be, since they differ by $0.24~\mathrm{km\,s^{-1}\,Mpc^{-1}}$ in $H_0$, and enlarging the catalogue tightens both equally rather than pulling them apart. LISA bright sirens are therefore a sharp arbiter of the Cepheid-calibrated scenario and, on this error budget, silent between the other two. We regard this asymmetry, rather than any ranking of the three, as the robust forecast.

Fourth, the gravitational-wave side supplies no prior of its own on the fiducial. Catalogues built at $H_0$ between $35$ and $115~\mathrm{km\,s^{-1}\,Mpc^{-1}}$ are recovered with fractional precision between $1.1\%$ and $2.3\%$: nothing in the waveform recognises an implausible expansion history, and what rejects the extreme fiducials is the electromagnetic comparison alone. The simulated data supply the error budget and the discriminating power that follows from it, but the fiducial remains an assumption of the exercise, which is why we quantify its effect explicitly instead of adopting a single choice.

Several limitations temper these conclusions. The headline tensions of Table~\ref{tab:tension} come from one noise realisation each; we quantify the resulting scatter at $\pm 0.4\sigma$ and it is what makes the \textit{Planck}18--Union3 comparison inconclusive, but the individual numbers should be read with that width attached. More seriously, the analysis assumes that every detected binary has an identified electromagnetic counterpart, an optimistic limit: realistic counterpart fractions for massive black hole binaries are considerably smaller, which would widen the LISA-only contours; the Fisher estimate of Section~\ref{sec:results:tiebreak} bounds the effect (the Pantheon+SH0ES exclusion exceeds $5\sigma$ at $N = 50$ and remains between $3$ and $5\sigma$ for $10$ to $20$ events), but a redshift-dependent counterpart fraction would be the proper treatment. Our merger population also follows a phenomenological rate and a single power law in mass rather than a semi-analytic galaxy-formation model, and it changes the component-mass range discontinuously at $z = 1$, a feature visible in Figs.~\ref{fig:budget} and~\ref{fig:population} that a physically motivated population would smooth. Semi-analytic models built on different black-hole seeding and delay prescriptions differ by more than an order of magnitude in predicted event count~\cite{klein2016science, barausse2020prospects}; what our population fixes is the distribution of per-event distance errors at a given $N$. Finally, the signal-to-noise integral uses a fixed upper frequency rather than the innermost stable circular orbit of each binary; we have checked that this affects $11\%$ of events and changes their median signal-to-noise ratio by $3\%$, well below the level at which any conclusion here would move.

These caveats define the natural next steps: folding in a redshift-dependent counterpart fraction, replacing the phenomenological population by a semi-analytic model with a continuous mass function, and averaging the reported tensions over many catalogue realisations rather than quoting a scatter. Extending the analysis beyond flat $\Lambda$CDM, to a dynamical dark energy equation of state or to modified GW propagation, would sharpen the role of LISA as an independent arbiter, and would do so in exactly the regime where the present analysis is informative: distinguishing histories that differ substantially, rather than adjudicating between neighbours.

\begin{acknowledgments}
MFAM is supported by the Coordenação de Aperfeiçoamento de Pessoal de Nível Superior (CAPES). LLS thanks the Paraíba State Research Foundation (FAPESQ) for financial support. This work used resources of the Centro Nacional de Processamento de Alto Desempenho em São Paulo (CENAPAD-SP).
\end{acknowledgments}

\bibliographystyle{JHEP}
\bibliography{referencias}

@article{sathyaprakash2009physics,
  title={Physics, astrophysics and cosmology with gravitational waves},
  author={Sathyaprakash, Bangalore Suryanarayana and Schutz, Bernard F},
  journal={Living reviews in relativity},
  volume={12},
  number={1},
  pages={2},
  year={2009},
  publisher={Springer}
}

@article{klein2016science,
  title={Science with the space-based interferometer eLISA: Supermassive black hole binaries},
  author={Klein, Antoine and Barausse, Enrico and Sesana, Alberto and Petiteau, Antoine and Berti, Emanuele and Babak, Stanislav and Gair, Jonathan and Aoudia, Sofiane and Hinder, Ian and Ohme, Frank and others},
  journal={Physical Review D},
  volume={93},
  number={2},
  pages={024003},
  year={2016},
  publisher={APS}
}

@article{wang2022forecast,
  title={Forecast for cosmological parameter estimation with gravitational-wave standard sirens from the LISA-Taiji network},
  author={Wang, Ling-Feng and Jin, Shang-Jie and Zhang, Jing-Fei and Zhang, Xin},
  journal={Science China Physics, Mechanics \& Astronomy},
  volume={65},
  number={1},
  pages={210411},
  year={2022},
  publisher={Springer}
}

@article{tamanini2016science,
  title={Science with the space-based interferometer eLISA. III: Probing the expansion of the Universe using gravitational wave standard sirens},
  author={Tamanini, Nicola and Caprini, Chiara and Barausse, Enrico and Sesana, Alberto and Klein, Antoine and Petiteau, Antoine},
  journal={Journal of Cosmology and Astroparticle Physics},
  volume={2016},
  number={04},
  pages={002},
  year={2016},
  publisher={IOP Publishing}
}

@article{aghanim2020planck,
  title={Planck 2018 results. VI. Cosmological parameters},
  author={{Planck Collaboration} and Aghanim, N. and Akrami, Y. and Ashdown, M. and Aumont, J. and Baccigalupi, C. and others},
  journal={Astronomy \& Astrophysics},
  volume={641},
  pages={A6},
  year={2020},
  eprint={1807.06209},
  archivePrefix={arXiv}
}

@article{abbott2023population,
  title={Population of merging compact binaries inferred using gravitational waves through GWTC-3},
  author={Abbott, Richard and Abbott, TD and Acernese, F and Ackley, K and Adams, C and Adhikari, N and Adhikari, RX and Adya, VB and Affeldt, C and Agarwal, D and others},
  journal={Physical Review X},
  volume={13},
  number={1},
  pages={011048},
  year={2023},
  publisher={APS}
}

@article{nishizawa2011tracing,
  title={Tracing the redshift evolution of Hubble parameter with gravitational-wave standard sirens},
  author={Nishizawa, Atsushi and Taruya, Atsushi and Saito, Shun},
  journal={Physical Review D—Particles, Fields, Gravitation, and Cosmology},
  volume={83},
  number={8},
  pages={084045},
  year={2011},
  publisher={APS}
}

@article{schneider2001low,
  title={Low-frequency gravitational waves from cosmological compact binaries},
  author={Schneider, Raffaella and Ferrari, Valeria and Matarrese, Sabino and Portegies Zwart, Simon F},
  journal={Monthly Notices of the Royal Astronomical Society},
  volume={324},
  number={4},
  pages={797--810},
  year={2001},
  publisher={Blackwell Science Ltd Oxford, UK}
}

@article{teixeira2023forecasts,
  title={Forecasts on interacting dark energy with standard sirens},
  author={Teixeira, Elsa M and Daniel, Richard and Frusciante, Noemi and van de Bruck, Carsten},
  journal={Physical Review D},
  volume={108},
  number={8},
  pages={084070},
  year={2023},
  publisher={APS}
}

@article{scolnic2022pantheon+,
  title={The Pantheon+ analysis: the full data set and light-curve release},
  author={Scolnic, Dan and Brout, Dillon and Carr, Anthony and Riess, Adam G and Davis, Tamara M and Dwomoh, Arianna and Jones, David O and Ali, Noor and Charvu, Pranav and Chen, Rebecca and others},
  journal={The Astrophysical Journal},
  volume={938},
  number={2},
  pages={113},
  year={2022},
  publisher={IOP Publishing}
}

@article{di2025cosmoverse,
  title={The CosmoVerse White Paper: Addressing observational tensions in cosmology with systematics and fundamental physics},
  author={Di Valentino, Eleonora and Said, Jackson Levi and Riess, Adam and Pollo, Agnieszka and Poulin, Vivian and G{\'o}mez-Valent, Adri{\`a} and Weltman, Amanda and Palmese, Antonella and Huang, Caroline D and van de Bruck, Carsten and others},
  journal={Physics of the Dark Universe},
  volume={49},
  pages={101965},
  year={2025},
  eprint={2504.01669},
  archivePrefix={arXiv}
}

@article{perivolaropoulos2022challenges,
  title={Challenges for $\Lambda$CDM: An update},
  author={Perivolaropoulos, Leandros and Skara, Foteini},
  journal={New Astronomy Reviews},
  volume={95},
  pages={101659},
  year={2022},
  publisher={Elsevier}
}

@article{holz2005using,
  title={Using gravitational-wave standard sirens},
  author={Holz, Daniel E and Hughes, Scott A},
  journal={The Astrophysical Journal},
  volume={629},
  number={1},
  pages={15},
  year={2005},
  publisher={IOP Publishing}
}

@article{schutz1986determining,
  title={Determining the Hubble constant from gravitational wave observations},
  author={Schutz, Bernard F},
  journal={Nature},
  volume={323},
  number={6086},
  pages={310--311},
  year={1986},
  publisher={Nature Publishing Group UK London}
}

@article{chen2018two,
  title={A two per cent Hubble constant measurement from standard sirens within five years},
  author={Chen, Hsin-Yu and Fishbach, Maya and Holz, Daniel E},
  journal={Nature},
  volume={562},
  number={7728},
  pages={545--547},
  year={2018},
  publisher={Nature Publishing Group UK London}
}

@article{auclair2023cosmology,
  title={Cosmology with the laser interferometer space antenna},
  author={Auclair, Pierre and Bacon, David and Baker, Tessa and Barreiro, Tiago and Bartolo, Nicola and Belgacem, Enis and Bellomo, Nicola and Ben-Dayan, Ido and Bertacca, Daniele and Besancon, Marc and others},
  journal={Living Reviews in Relativity},
  volume={26},
  number={1},
  pages={5},
  year={2023},
  publisher={Springer}
}

@article{arun2022new,
  title={New horizons for fundamental physics with LISA},
  author={Arun, KG and Belgacem, Enis and Benkel, Robert and Bernard, Laura and Berti, Emanuele and Bertone, Gianfranco and Besancon, Marc and Blas, Diego and B{\"o}hmer, Christian G and Brito, Richard and others},
  journal={Living Reviews in Relativity},
  volume={25},
  number={1},
  pages={4},
  year={2022},
  publisher={Springer}
}

@article{riess2022comprehensive,
  title={A comprehensive measurement of the local value of the Hubble constant with 1 km s- 1 Mpc- 1 uncertainty from the Hubble Space Telescope and the SH0ES team},
  author={Riess, Adam G and Yuan, Wenlong and Macri, Lucas M and Scolnic, Dan and Brout, Dillon and Casertano, Stefano and Jones, David O and Murakami, Yukei and Anand, Gagandeep S and Breuval, Louise and others},
  journal={The Astrophysical journal letters},
  volume={934},
  number={1},
  pages={L7},
  year={2022},
  publisher={The American Astronomical Society}
}

@article{di2021realm,
  title={In the realm of the Hubble tension—a review of solutions},
  author={Di Valentino, Eleonora and Mena, Olga and Pan, Supriya and Visinelli, Luca and Yang, Weiqiang and Melchiorri, Alessandro and Mota, David F and Riess, Adam G and Silk, Joseph},
  journal={Classical and Quantum Gravity},
  volume={38},
  number={15},
  pages={153001},
  year={2021},
  publisher={IOP Publishing}
}

@article{perivolaropoulos2024hubble,
  title={Hubble tension or distance ladder crisis?},
  author={Perivolaropoulos, Leandros},
  journal={Physical Review D},
  volume={110},
  number={12},
  pages={123518},
  year={2024},
  publisher={APS}
}

@article{freedman2025status,
  title={Status report on the Chicago-Carnegie Hubble Program (CCHP): measurement of the Hubble constant using the Hubble and James Webb space telescopes},
  author={Freedman, Wendy L and Madore, Barry F and Hoyt, Taylor J and Jang, In Sung and Lee, Abigail J and Owens, Kayla A},
  journal={The Astrophysical Journal},
  volume={985},
  number={2},
  pages={203},
  year={2025},
  publisher={The American Astronomical Society}
}

@article{riess2024jwst,
  title={JWST validates HST distance measurements: Selection of supernova subsample explains differences in JWST estimates of local H 0},
  author={Riess, Adam G and Scolnic, Dan and Anand, Gagandeep S and Breuval, Louise and Casertano, Stefano and Macri, Lucas M and Li, Siyang and Yuan, Wenlong and Huang, Caroline D and Jha, Saurabh and others},
  journal={The Astrophysical Journal},
  volume={977},
  number={1},
  pages={120},
  year={2024},
  publisher={The American Astronomical Society}
}

@article{riess2024jwst2,
  title={JWST observations reject unrecognized crowding of Cepheid photometry as an explanation for the Hubble tension at 8$\sigma$ confidence},
  author={Riess, Adam G and Anand, Gagandeep S and Yuan, Wenlong and Casertano, Stefano and Dolphin, Andrew and Macri, Lucas M and Breuval, Louise and Scolnic, Dan and Perrin, Marshall and Anderson, Richard I},
  journal={The Astrophysical Journal Letters},
  volume={962},
  number={1},
  pages={L17},
  year={2024},
  publisher={IOP Publishing}
}

@article{abbott2017gw170817,
  title={GW170817: observation of gravitational waves from a binary neutron star inspiral},
  author={Abbott, Benjamin P and Abbott, Rich and Abbott, Thomas D and Acernese, Fausto and Ackley, Kendall and Adams, Carl and Adams, Thomas and Addesso, Paolo and Adhikari, Rana X and Adya, Vaishali B and others},
  journal={Physical review letters},
  volume={119},
  number={16},
  pages={161101},
  year={2017},
  publisher={APS}
}

@article{fishbach2019standard,
  title={A standard siren measurement of the Hubble constant from GW170817 without the electromagnetic counterpart},
  author={Fishbach, Maya and Gray, R and Maga{\~n}a Hernandez, I and Qi, H and Sur, A and Acernese, F and Aiello, L and Allocca, A and Aloy, MA and Amato, A and others},
  journal={The Astrophysical Journal Letters},
  volume={871},
  number={1},
  pages={L13},
  year={2019},
  publisher={The American Astronomical Society}
}

@article{soares2019first,
  title={First measurement of the Hubble constant from a dark standard siren using the dark energy survey galaxies and the LIGO/Virgo binary--black-hole merger GW170814},
  author={Soares-Santos, Marcelle and Palmese, Antonella and Hartley, W and Annis, James and Garcia-Bellido, J and Lahav, O and Doctor, Z and Fishbach, M and Holz, DE and Lin, H and others},
  journal={The Astrophysical Journal Letters},
  volume={876},
  number={1},
  pages={L7},
  year={2019},
  publisher={The American Astronomical Society}
}

@article{mastrogiovanni2024cosmology,
  title={Cosmology with gravitational waves: a review},
  author={Mastrogiovanni, Simone and Karathanasis, Christos and Gair, Jonathan and Ashton, Gregory and Rinaldi, Stefano and Huang, Hsiang-Yu and Dalya, Gergely},
  journal={Annalen der Physik},
  volume={536},
  number={2},
  pages={2200180},
  year={2024},
  publisher={Wiley Online Library}
}

@article{amaro2017laser,
  title={Laser interferometer space antenna},
  author={Amaro-Seoane, Pau and Audley, Heather and Babak, Stanislav and Baker, John and Barausse, Enrico and Bender, Peter and Berti, Emanuele and Binetruy, Pierre and Born, Michael and Bortoluzzi, Daniele and others},
  journal={arXiv preprint arXiv:1702.00786},
  note={A proposal in response to the ESA call for L3 mission concepts},
  year={2017}
}

@article{barausse2020prospects,
  title={Prospects for fundamental physics with LISA},
  author={Barausse, Enrico and Berti, Emanuele and Hertog, Thomas and Hughes, Scott A and Jetzer, Philippe and Pani, Paolo and Sotiriou, Thomas P and Tamanini, Nicola and Witek, Helvi and Yagi, Kent and others},
  journal={General Relativity and Gravitation},
  volume={52},
  number={8},
  pages={81},
  year={2020},
  publisher={Springer}
}

@article{speri2021testing,
  title={Testing the quasar Hubble diagram with LISA standard sirens},
  author={Speri, Lorenzo and Tamanini, Nicola and Caldwell, Robert R and Gair, Jonathan R and Wang, Benjamin},
  journal={Physical Review D},
  volume={103},
  number={8},
  pages={083526},
  year={2021},
  publisher={APS}
}

@article{mukherjee2021velocity,
  title={Velocity correction for Hubble constant measurements from standard sirens},
  author={Mukherjee, Suvodip and Lavaux, Guilhem and Bouchet, Fran{\c{c}}ois R and Jasche, Jens and Wandelt, Benjamin D and Nissanke, Samaya and Leclercq, Florent and Hotokezaka, Kenta},
  journal={Astronomy \& Astrophysics},
  volume={646},
  pages={A65},
  year={2021},
  publisher={EDP Sciences}
}

@article{belgacem2019testing,
  title={Testing modified gravity at cosmological distances with LISA standard sirens},
  author={Belgacem, Enis and Calcagni, Gianluca and Crisostomi, Marco and Dalang, Charles and Dirian, Yves and Ezquiaga, Jose Mar{\'\i}a and Fasiello, Matteo and Foffa, Stefano and Ganz, Alexander and Garc{\'\i}a-Bellido, Juan and others},
  journal={Journal of Cosmology and Astroparticle Physics},
  volume={2019},
  number={07},
  pages={024--024},
  year={2019}
}

@article{mangiagli2024massive,
  title={Massive black hole binaries in LISA: Constraining cosmological parameters at high redshifts},
  author={Mangiagli, Alberto and Caprini, Chiara and Marsat, Sylvain and Speri, Lorenzo and Caldwell, Robert R and Tamanini, Nicola},
  journal={Physical Review D},
  volume={111},
  number={8},
  pages={083043},
  year={2025},
  publisher={APS}
}

@article{rubin2025union3,
  title={Union through UNITY: cosmology with 2000 SNe using a unified Bayesian framework},
  author={Rubin, David and Aldering, Greg and Betoule, Marc and Fruchter, Andy and Huang, Xiaosheng and Kim, Alex G and Lidman, Chris and Linder, Eric and Perlmutter, Saul and Ruiz-Lapuente, Pilar and others},
  journal={The Astrophysical Journal},
  volume={986},
  number={2},
  pages={231},
  year={2025},
  publisher={The American Astronomical Society}
}

@article{robson2019construction,
  title={The construction and use of LISA sensitivity curves},
  author={Robson, Travis and Cornish, Neil J and Liu, Chang},
  journal={Classical and Quantum Gravity},
  volume={36},
  number={10},
  pages={105011},
  year={2019},
  publisher={IOP Publishing}
}

@article{cornish2003lisa,
  title={LISA response function},
  author={Cornish, Neil J and Rubbo, Louis J},
  journal={Physical Review D},
  volume={67},
  number={2},
  pages={022001},
  year={2003},
  publisher={APS}
}

@article{moresco2022unveiling,
  title={Unveiling the Universe with emerging cosmological probes},
  author={Moresco, Michele and Amati, Lorenzo and Amendola, Luca and Birrer, Simon and Blakeslee, John P and Cantiello, Michele and Cimatti, Andrea and Darling, Jeremy and Della Valle, Massimo and Fishbach, Maya and others},
  journal={Living Reviews in Relativity},
  volume={25},
  number={1},
  pages={6},
  year={2022},
  publisher={Springer}
}

@article{foreman2013emcee,
  title={emcee: the MCMC hammer},
  author={Foreman-Mackey, Daniel and Hogg, David W and Lang, Dustin and Goodman, Jonathan},
  journal={Publications of the Astronomical Society of the Pacific},
  volume={125},
  number={925},
  pages={306--312},
  year={2013},
  publisher={University of Chicago Press}
}

@article{desi2025dr2,
  title={DESI DR2 results. II. Measurements of baryon acoustic oscillations and cosmological constraints},
  author={{DESI Collaboration}},
  journal={Physical Review D},
  volume={112},
  number={8},
  pages={083515},
  year={2025},
  publisher={APS},
  eprint={2503.14738},
  archivePrefix={arXiv},
  primaryClass={astro-ph.CO}
}

@article{brout2022pantheon,
  title={The Pantheon+ analysis: cosmological constraints},
  author={Brout, Dillon and Scolnic, Dan and Popovic, Brodie and Riess, Adam G and Carr, Anthony and Zuntz, Joe and Kessler, Rick and Davis, Tamara M and Hinton, Samuel and Jones, David and others},
  journal={The Astrophysical Journal},
  volume={938},
  number={2},
  pages={110},
  year={2022},
  publisher={IOP Publishing}
}

@article{chen2019distance,
  title={Distance priors from Planck final release},
  author={Chen, Lu and Huang, Qing-Guo and Wang, Ke},
  journal={Journal of Cosmology and Astroparticle Physics},
  volume={2019},
  number={02},
  pages={028},
  year={2019},
  publisher={IOP Publishing}
}

@article{kocsis2006finding,
  title={Finding the electromagnetic counterparts of cosmological standard sirens},
  author={Kocsis, Bence and Frei, Zsolt and Haiman, Zolt{\'a}n and Menou, Kristen},
  journal={The Astrophysical Journal},
  volume={637},
  number={1},
  pages={27--37},
  year={2006},
  publisher={IOP Publishing},
  eprint={astro-ph/0505394},
  archivePrefix={arXiv}
}

@article{brieden2023model,
  title={A tale of two (or more) h's},
  author={Brieden, Samuel and Gil-Mar{\'\i}n, H{\'e}ctor and Verde, Licia},
  journal={Journal of Cosmology and Astroparticle Physics},
  volume={2023},
  number={04},
  pages={023},
  year={2023},
  publisher={IOP Publishing},
  eprint={2212.04522},
  archivePrefix={arXiv}
}

@article{schoneberg2024bbn,
  title={The 2024 BBN baryon abundance update},
  author={Sch{\"o}neberg, Nils},
  journal={Journal of Cosmology and Astroparticle Physics},
  volume={2024},
  number={06},
  pages={006},
  year={2024},
  publisher={IOP Publishing},
  eprint={2401.15054},
  archivePrefix={arXiv}
}

@article{raveri2019concordance,
  title={Concordance and discordance in cosmology},
  author={Raveri, Marco and Hu, Wayne},
  journal={Physical Review D},
  volume={99},
  number={4},
  pages={043506},
  year={2019},
  publisher={American Physical Society}
}

@article{goliath2001supernovae,
  title={Supernovae and the nature of the dark energy},
  author={Goliath, M and Amanullah, R and Astier, P and Goobar, A and Pain, R},
  journal={Astronomy \& Astrophysics},
  volume={380},
  number={1},
  pages={6--18},
  year={2001},
  publisher={EDP Sciences}
}

@article{popovic2025dovekie,
  title={The Dark Energy Survey Supernova Program: A reanalysis of cosmology results and evidence for evolving dark energy with an updated Type Ia supernova calibration},
  author={Popovic, B. and Shah, P. and Kenworthy, W. D. and Kessler, R. and Davis, T. M. and others},
  journal={Monthly Notices of the Royal Astronomical Society},
  volume={548},
  number={4},
  pages={stag632},
  year={2026},
  eprint={2511.07517},
  archivePrefix={arXiv}
}

@article{goodman2010ensemble,
  title={Ensemble samplers with affine invariance},
  author={Goodman, Jonathan and Weare, Jonathan},
  journal={Communications in Applied Mathematics and Computational Science},
  volume={5},
  number={1},
  pages={65--80},
  year={2010},
  publisher={Mathematical Sciences Publishers}
}

@article{gelman1992inference,
  title={Inference from iterative simulation using multiple sequences},
  author={Gelman, Andrew and Rubin, Donald B},
  journal={Statistical Science},
  volume={7},
  number={4},
  pages={457--472},
  year={1992}
}

@article{hu1996small,
  title={Small-scale cosmological perturbations: an analytic approach},
  author={Hu, Wayne and Sugiyama, Naoshi},
  journal={The Astrophysical Journal},
  volume={471},
  pages={542},
  year={1996}
}

@article{moresco2012improved,
  title={Improved constraints on the expansion rate of the Universe up to z$\sim$1.1 from the spectroscopic evolution of cosmic chronometers},
  author={Moresco, M. and Cimatti, A. and Jimenez, R. and Pozzetti, L. and Zamorani, G. and others},
  journal={Journal of Cosmology and Astroparticle Physics},
  volume={2012},
  number={08},
  pages={006},
  year={2012},
  eprint={1201.3609},
  archivePrefix={arXiv}
}

@article{moresco2015raising,
  title={Raising the bar: new constraints on the Hubble parameter with cosmic chronometers at z$\sim$2},
  author={Moresco, Michele},
  journal={Monthly Notices of the Royal Astronomical Society},
  volume={450},
  number={1},
  pages={L16--L20},
  year={2015},
  eprint={1503.01116},
  archivePrefix={arXiv}
}

@article{moresco2016sixpercent,
  title={A 6\% measurement of the Hubble parameter at z$\sim$0.45: direct evidence of the epoch of cosmic re-acceleration},
  author={Moresco, M. and Pozzetti, L. and Cimatti, A. and Jimenez, R. and Maraston, C. and others},
  journal={Journal of Cosmology and Astroparticle Physics},
  volume={2016},
  number={05},
  pages={014},
  year={2016},
  eprint={1601.01701},
  archivePrefix={arXiv}
}

@article{moresco2020setting,
  title={Setting the stage for cosmic chronometers. II. Impact of stellar population synthesis models systematics and full covariance matrix},
  author={Moresco, Michele and Jimenez, Raul and Verde, Licia and Cimatti, Andrea and Pozzetti, Lucia},
  journal={The Astrophysical Journal},
  volume={898},
  number={1},
  pages={82},
  year={2020},
  eprint={2003.07362},
  archivePrefix={arXiv}
}

@article{abbott2017gravitational,
  title={A gravitational-wave standard siren measurement of the Hubble constant},
  author={{LIGO Scientific Collaboration and Virgo Collaboration} and others},
  journal={Nature},
  volume={551},
  pages={85--88},
  year={2017},
  eprint={1710.05835},
  archivePrefix={arXiv}
}

@article{abbott2023constraints,
  title={Constraints on the cosmic expansion history from GWTC-3},
  author={{LIGO Scientific Collaboration and Virgo Collaboration and KAGRA Collaboration} and others},
  journal={The Astrophysical Journal},
  volume={949},
  number={2},
  pages={76},
  year={2023},
  eprint={2111.03604},
  archivePrefix={arXiv}
}

@article{zhao2011determination,
  title={Determination of dark energy by the Einstein Telescope: Comparing with CMB, BAO, and SNIa observations},
  author={Zhao, Wen and Van Den Broeck, Chris and Baskaran, Deepak and Li, Tjonnie G. F.},
  journal={Physical Review D},
  volume={83},
  number={2},
  pages={023005},
  year={2011},
  eprint={1009.0206},
  archivePrefix={arXiv}
}

@article{cai2017estimating,
  title={Estimating cosmological parameters by the simulated data of gravitational waves from the Einstein Telescope},
  author={Cai, Rong-Gen and Yang, Tao},
  journal={Physical Review D},
  volume={95},
  number={4},
  pages={044024},
  year={2017},
  eprint={1608.08008},
  archivePrefix={arXiv}
}

@article{hirata2010reducing,
  title={Reducing the weak lensing noise for the gravitational wave Hubble diagram using the non-Gaussianity of the magnification distribution},
  author={Hirata, Christopher M. and Holz, Daniel E. and Cutler, Curt},
  journal={Physical Review D},
  volume={81},
  number={12},
  pages={124046},
  year={2010},
  eprint={1004.3988},
  archivePrefix={arXiv}
}

@article{cutler2006bbo,
  title={Big Bang Observer and the neutron-star-binary subtraction problem},
  author={Cutler, Curt and Harms, Jan},
  journal={Physical Review D},
  volume={73},
  number={4},
  pages={042001},
  year={2006},
  eprint={gr-qc/0511092},
  archivePrefix={arXiv}
}

@article{cutler1998angular,
  title={Angular resolution of the LISA gravitational wave detector},
  author={Cutler, Curt},
  journal={Physical Review D},
  volume={57},
  number={12},
  pages={7089--7102},
  year={1998},
  eprint={gr-qc/9703068},
  archivePrefix={arXiv}
}

@article{mangiagli2022massive,
  title={Massive black hole binaries in LISA: multimessenger prospects and electromagnetic counterparts},
  author={Mangiagli, Alberto and Caprini, Chiara and Volonteri, Marta and Marsat, Sylvain and Vergani, Susanna and Tamanini, Nicola and Inchausp{\'e}, Henri},
  journal={Physical Review D},
  volume={106},
  number={10},
  pages={103017},
  year={2022},
  eprint={2207.10678},
  archivePrefix={arXiv}
}
\end{document}